\documentclass[12pt]{article}
\usepackage[a4paper, total={6.5in, 9in}, left=1in, right=1in]{geometry}
\usepackage{amsmath}
\usepackage{caption}
\usepackage{subcaption}
\usepackage{graphicx}
\usepackage{booktabs}
\usepackage{enumerate}
\usepackage[comma]{natbib}
\usepackage{url}
\usepackage{siunitx}
\usepackage{amsthm}
\usepackage[linesnumbered,ruled,vlined]{algorithm2e}
\usepackage{amsmath,amsfonts,accents,url,comment,float,multirow,multicol,adjustbox,makecell,appendix,dsfont}
\usepackage{bm}
\usepackage{lineno}

\newcommand\independent{\protect\mathpalette{\protect\independenT}{\perp}}
\def\independenT#1#2{\mathrel{\rlap{$#1#2$}\mkern2mu{#1#2}}}

\newtheorem{theorem}{Theorem}
\newtheorem{assumption}{Assumption}

\usepackage{xcolor}
\usepackage{hyperref}
\usepackage{amsmath}
\usepackage{cleveref} 
\hypersetup{
    colorlinks=true,
    linkcolor=red,      
    citecolor=green,    
    urlcolor=blue       
}

\newif\ifshowupdates
\showupdatestrue      

\title{
Multivariate Low-Rank State-Space Model with SPDE Approach for High-Dimensional Data}

\author{
  Jacopo Rodeschini\thanks{Department of Engineering and Applied Sciences, University of Bergamo, Dalmine, Italy. Email: \href{mailto:jacopo.rodeschini@unibg.it}{jacopo.rodeschini@unibg.it}} 
  \and 
  Lorenzo Tedesco\thanks{Department of Economics, University of Bergamo, Bergamo, Italy}  
  \and
  Francesco Finazzi\footnotemark[2]
  \and
  Philipp Otto\thanks{School of Mathematics and Statistics, University of Glasgow, Glasgow, UK} 
  \and 
  Alessandro Fassò\footnotemark[2]
}

\date{}
\begin{document}
	
\maketitle
	
	\begin{abstract}
		This paper proposes a novel low-rank approximation to the multivariate State-Space Model. 
		The Stochastic Partial Differential Equation (SPDE) approach is applied component-wise to the independent-in-time Matérn Gaussian innovation term in the latent equation, assuming component independence. This results in a sparse representation of the latent process on a finite element mesh, allowing for scalable inference through sparse matrix operations. Dependencies among observed components are introduced through a matrix of weights applied to the latent process. Model parameters are estimated using the Expectation-Maximisation algorithm, which features closed-form updates for most parameters and efficient numerical routines for the remaining parameters. We prove theoretical results regarding the accuracy and convergence of the SPDE-based approximation under fixed-domain asymptotics. Simulation studies show our theoretical results. We include an empirical application on air quality to demonstrate the practical usefulness of the proposed model, which maintains computational efficiency in high-dimensional settings. In this application, we reduce computation time by about 93\%, with only a 15\% increase in the validation error.
	\end{abstract}
	\noindent%
	\vfill
	
	\newpage
	


\section{Introduction}\label{sec:introduction}
Spatio-temporal statistics is a discipline that focuses on analysing data indexed by both space and time \cite{cressie2011statistics}, and has become increasingly popular in geosciences due to extensive sampling campaigns, growing data availability, and a wide range of applications, including environmental and climate sciences. 

Often, multiple variables associated with space and time are observed together. The observed values of these spatio-temporal variables are referred to as multivariate spatio-temporal data, and typically exhibit three forms of dependence. Temporal dependence reflects the causal nature of time-evolving processes, where past observations influence future outcomes. Spatial dependence, on the other hand, is generally symmetric and instantaneous; nearby locations tend to be similar, while similarity usually diminishes with distance. A third layer of complexity arises from cross-variable dependence, where correlations exist between different variables, either at the same location or across distinct spatial sites. The most informative scenario is the isotopic configuration, in which all variables are measured at every site. In contrast, the completely heterotopic configuration refers to the case where each variable is observed only at distinct, non-overlapping locations. A less restricted case is the partially heterotopic setting, in which only some variables share some sites. Formal definitions of isotopy and heterotopy are found in \cite{wackernagel2003multivariate}.

Modelling both spatial and temporal dependencies is essential for accurate prediction and interpretation \citep{zhang2007maximum}. To address this, multivariate spatio-temporal models for heterotopic data have been developed to capture these intricate interdependencies \citep{finazzi2013model}. Such models enhance our ability to uncover underlying processes, improve predictive accuracy, and support more effective policy-making and management decisions.


Among the various modelling approaches, Gaussian Fields (GFs) have become a widely used and powerful tool for solving regression problems in spatial and spatio-temporal data \citep{rasmussen2003gaussian}. Their popularity stems from their flexibility and their natural ability to capture spatial and temporal dependencies \citep{stein1999interpolation}. The underlying regression function is assumed to be a realisation of a GF, whose mean and covariance function must be evaluated. Then, using the conditional multivariate normal distribution, a GF provides closed-form prediction as well as a predictive interval.



With the widespread availability of remote sensing platforms, especially satellite-based systems, the volume and resolution of spatial data have increased dramatically, see for instance \cite{datta2016hierarchical, heaton2019case}, thereby significantly amplifying the computational demands associated with GF modelling. This challenge is known as the ``Big $n$ problem''\citep{jona2013discussing}: the computational cost of inverting a covariance matrix scales poorly with the number of spatio-temporal observations $n = mT$, where $m$ is the total number of locations and $T$ is the number of time steps, since covariance matrices grow with complexity $\mathcal{O}(m^2T^2)$ in storage and $\mathcal{O}(m^3T^3)$ in inversion. These demands quickly exceed available resources under increasing data volume, rendering exact inference infeasible even with a moderately large $n$. 

In the context of GFs, a range of strategies has been developed to address the high computational cost associated with large datasets. Comprehensive reviews of these methods can be found in \cite{heaton2019case} and \cite{liu2020gaussian}. Three major classes of approximation techniques are commonly used: likelihood approximations, low-rank approximations, and sparse representations. A key drawback of using computationally efficient approximated models is the need to assess how closely the approximate predictors and their mean squared prediction errors match the true, theoretical values.

Likelihood approximations, such as the Vecchia approximation \citep{vecchia1988estimation, katzfuss2020vecchia, katzfuss2021general}, reformulate the joint distribution of observations into a product of conditional distributions, where each observation conditions only on a subset of others. This reduction improves computational efficiency, but the choice of conditioning sets is critical. Covariance tapering for univariate \citep{furrer2006covariance, kaufman2008covariance} and multivariate \citep{bevilacqua2016covariance} process is another likelihood-based method that imposes sparsity in the covariance by multiplying the original covariance function by a compactly supported taper function. However, tapering assumes that spatial dependence vanishes beyond a fixed distance, which may underestimate long-range correlations \citep{stein1999interpolation,zhang2004inconsistent}. Low-rank approximations reduce dimensionality by expressing the GF as a linear combination of a small number $R \ll m$ of basis functions. Lowering the rank of the covariance matrix significantly cuts computational cost. Well-known example include Fixed-Rank Kriging \citep{cressie2008fixed, zammit2021frk} and LatticeKrig \citep{nychka2015multiresolution}. However, the choice of basis functions involves multiple tuning parameters, such as the types of basis functions, levels of resolution, and information in the coarsest resolution, which can lead to issues with identifiability and overfitting \citep{song2024large}.

A distinct and powerful class of approximations of spatial GFs is based on Gaussian Markov Random Fields (GMRFs) \citep{rue2005gaussian}, where conditional independence across a graph structure yields a sparse precision matrix. Notably, \cite{lindgren2011explicit} established a link between Matérn GFs and GMRFs by solving a Stochastic Partial Differential Equation (SPDE) using the Finite Element Method (FEM). This SPDE approach enables linear-cost computation of the precision matrix (inverse of the covariance matrix) over complex domains and has recently been extended to spatio-temporal processes \citep{clarotto2024spde}.

The State-Space Model (SSM) is a widely used framework for formulating Gaussian regression problems \citep{shumway2000time}, particularly in spatio-temporal contexts. It explicitly models one-directional temporal dependence through a latent process. Given fixed parameters, inference on the unobserved latent states can be efficiently performed using the Kalman filter, leveraging the separability of the covariance structure \citep{hamilton2020time}. In the more common case where model parameters are unknown, the Expectation-Maximisation (EM) algorithm is typically employed for parameter estimation \cite{shumway2000time}. Due to these advantages, SSM have been extensively used in spatio-temporal modelling; see, for instance, \cite{xu2007estimation,cressie2002space,fasso2011maximum} and \cite{calculli2015maximum} for extensions to heterotopic data. 

While inference using the Kalman filter scales linearly with time, $\mathcal{O}(T)$, it incurs a cubic cost in the number of spatial locations, $\mathcal{O}(m^3)$, which becomes prohibitive for large $m$. To address this issue, several approximation strategies have been proposed. For instance, covariance tapering has been applied to SSMs to induce sparsity in the spatial structure \citep{bevilacqua2010weighted}. Alternatively, low-rank approximations, such as those based on FRK, have been applied within the SSM framework \citep{cressie2010fixed}. More recently, \cite{schmidt2023rank} proposed a method that reduces rank via a truncated singular value decomposition of the conditional covariance matrix, though this may come at the cost of losing physical interpretability of the latent process.

In this paper, we introduce a novel multivariate Low-Rank State-Space Model (LR-SSM) designed for heterotopic observations. Our approach incorporates the SPDE framework to significantly reduce the rank of the spatial covariance matrix, from $m$ to $ R \ll m$. This reduces the computational cost of the Kalman filter to $\mathcal{O}(mR^2T)$, while preserving the physical interpretation of the original model. We prove theoretical results, including asymptotic properties and error bounds for the proposed low-rank model. The maximum likelihood estimation (MLE) of the model parameters is carried out via EM, which results in closed-form updates for most parameters. Extensive simulations validate both the accuracy and efficiency of the method. Although our discussion focuses on domains $\Omega \subset \mathbb{R}^2$, the approach extends naturally to higher-dimensional manifolds, and future studies may further adapt it to networks and graphs, highlighting the model’s flexibility and broad applicability.

The remainder of this article is organised as follows. Section \ref{sec:ssm_framework} introduces the formulation of the Gaussian regression problem using the SSM framework and presents the equations defining the proposed LR-SSM, tailored for large-scale multivariate spatio-temporal datasets. We also provide theoretical results and examine how the low-rank approximation affects inference. In Section \ref{sec:estimation}, we detail the ML estimation procedure, with a focus on the low-rank representation derived via the SPDE approach. Section \ref{sec:simulation} presents numerical studies that illustrate the performance of our method and confirm the theoretical results. In Section \ref{sec:case_study}, we present the application of the LR-SSM to bivariate air quality data. Finally, Section \ref{sec:conclusion} concludes the paper and outlines future development.

\section{Multivariate State-Space Model Framework}\label{sec:ssm_framework}
In this section, we present the notation for the multivariate SSM framework. Let $\bm{y}(\bm{s}, t) = (y_1(\bm{s}, t), \dots, y_p(\bm{s}, t))' \in \mathbb{R}^p$ be a $p$-dimensional spatio-temporal random field defined over a spatial domain $\Omega \subset \mathbb{R}^2$ and discrete time steps $t = 1, \dots, T$, where $\bm{s} \in \Omega$ indexes the spatial location.  We consider the classical spatio-temporal random effects model \citep{zammit2021frk}, expressing $\bm{y}(\bm{s}, t)$ through the measurement equation:
\begin{align}
\label{eq:state_space1}
\bm{y}(\bm{s}, t) = \bm{X}'(\bm{s}, t)\bm{\beta} + \bm{W}\bm{z}(\bm{s}, t) + \bm{\varepsilon}(\bm{s}, t),
\end{align}
where $ \bm{X}(\bm{s}, t) = \operatorname{blockdiag}(\bm{x}_1(\bm{s}, t), \dots, \bm{x}_p(\bm{s}, t)) $ is a $b \times p$ design matrix,  $\bm{x}_i(\bm{s}, t)$ is a column vector of dimension $b_i$ associated with $y_i(\bm{s}, t)$ for $i = 1,\dots p$, and $b = \sum_{i=1}^p b_i$. The parameter $\bm{\beta} = (\bm \beta_1', \dots, \bm \beta_p')'$ is a $b$-dimensional vector of regression coefficients, where each $\bm \beta_i$, for $i = 1, \dots, p$, is a column vector of dimension $b_i$ corresponding to the regression covariates associated with $y_i(\bm{s}, t)$. The operator $\operatorname{blockdiag}(\cdot)$ is the block diagonal matrix operator. 
$\bm{W}$ is a $p \times q$ loading matrix that linearly projects the latent random effects $\bm{z}(\bm{s}, t)$ into the observation space. The term $\bm{\varepsilon}(\bm{s}, t)$ represents the Gaussian measurement error, assumed to be statistically independent across components, space and time.  We denote by  $\bm \Sigma=\operatorname{diag}(\bm \sigma^2)$ the diagonal variance matrix of $\bm \varepsilon(\bm s, t)$, with $\bm \sigma^2 = (\sigma_1^2, \dots, \sigma_p^2)'\in \mathbb{R}^p_+$ being the vector of measurement error variances. 
The latent process $\bm{z}(\bm{s}, t)$ evolves over time, following the state equation:
\begin{align}
\label{eq:state_space2}
\bm{z}(\bm{s}, t) = \bm{F}\bm{z}(\bm{s}, t-1) + \bm{\eta}(\bm{s}, t),
\end{align}

\noindent where $\bm{F} = \operatorname{diag}(\bm{f})$ denotes a $q \times q$ diagonal transition matrix describing the system’s dynamics, and $\bm{f} = (f_1, \dots, f_q)'$ with each $f_i \in (-1, 1)$ to ensure the stationarity of $\bm{z}(\bm{s}, t)$. Moreover, for every $\bm s\in\Omega$, $\bm z(\bm s, 0)$ is an independent Gaussian variable with mean $\bm \mu_0\in \mathbb{R}^q$ and $q\times q$ covariance matrix $\bm \Sigma_0$. The term $\bm{\eta}(\bm{s}, t) = (\eta_1(\bm{s}, t), \dots, \eta_q(\bm{s}, t))'$ represents a $q$-dimensional, zero-mean Gaussian innovation process which is temporally independent. Moreover, we assume that the components of $\bm{\eta}(\bm{s}, t)$ are mutually independent, i.e., $\eta_j(\bm{s}, t) \independent \eta_{j'}(\bm{s}', t')$ for all $j \ne j'$, at any spatial locations $\bm{s}, \bm{s}'$ and times $t, t'$. We also assume that each $\eta_i(\bm s,t)$ is a zero-mean Màtern GF, with rescale parameter $\kappa_i$, unit marginal variance and smoothness parameter $\nu_i$. Specifically, the Matérn covariance function between two spatial locations $\bm{s}$ and $\bm{s}'$ is given by
\begin{equation}\label{eq:matern}
\text{Cov}(\eta_i(\bm{s},t), \eta_i(\bm{s}',t)) = \frac{1}{\Gamma(\nu_i) 2^{\nu_i - 1}} (\kappa_i \|\bm{s} - \bm{s}'\|)^{\nu_i} K_{\nu_i}(\kappa_i \|\bm{s} - \bm{s}'\|),
\end{equation}
where $K_{\nu_i}$ is the modified Bessel function of the second kind, and $\kappa_i$ controls the spatial range of dependence. The smoothness parameter is rarely estimated directly; instead, it is typically set a priori based on the desired level of smoothness in the Matérn process or informed by physical assumptions about the underlying phenomenon. In what follows, we set $\nu_i = 1$, which corresponds to a spatial process that is continuous but not mean-square differentiable. This choice strikes a balance between model flexibility and computational efficiency, making it well-suited for applications where a certain degree of spatial roughness is expected or acceptable. While the proposed methodology allows for alternative values of $\nu_i$, we adopt this setting to streamline both the notation and the computational workload.  


Together, the measurement and state equations \eqref{eq:state_space1}--\eqref{eq:state_space2} define the linear SSM \citep{shumway2000time}. This formulation is particularly appealing due to its generality and flexibility, and is capable of capturing both observed and latent components of the time series. 
The model parameter set that needs to be estimated is:
\begin{equation} \label{eq:Pi}
\Pi = \{\bm{\beta}, \bm{\sigma}^2, \bm{f}, \bm W, k_1, \dots, k_q\},
\end{equation} 

As $\bm{\varepsilon}(\bm{s}, t)$ and $\bm{\eta} (\bm{s}, t)$ are Gaussian, the inference on $\bm{z}(\bm{s}, t)$ is carried out efficiently using the Kalman filter \citep{hamilton2020time} which results in the best linear estimator in the minimum mean-square-error sense \citep{HumpherysFresh}. As we discussed in the introduction, however, the Kalman filter requires the inversion of an $m\times m$ matrix at every time $t = 1,\dots, T$, which results in a cost of $\mathcal{O}(m^3T)$. While linear in time, it does not scale properly in terms of the number of spatial locations $m$. 

\subsection{Multivariate Low-Rank SPDE State-Space Model}
\label{sec:model}
In this section, we introduce a low-rank SSM based on the SPDE approach \citep{lindgren2011explicit}, designed to approximate the data-generating process \eqref{eq:state_space1}--\eqref{eq:state_space2} and to enable inference on the latent dynamics at a reduced computational cost of $\mathcal{O}(mR^2T)$ with $R \ll m$. We also examine the special case $R = m$. To this end, we discuss how to replace the GF $\eta_i$ in Equation~\eqref{eq:state_space2} with a low-rank approximation by expressing $\eta_i$ as a linear combination of $R$ finite basis functions, yielding a finite-dimensional representation of the state equation and, ultimately, computational gains.

\citet{whittle1954stationary, whittle1963stochastic} showed that a Matérn GF can be characterised as the unique solution to a specific stochastic partial differential equation (SPDE) defined over the domain $\Omega$. By applying the Galerkin method \citep{ern2004theory} to this SPDE, one obtains a low-rank approximation of the field $\eta_i$, which corresponds to a Gaussian Markov Random Field (GMRF) representation of the original GF. \citet{lindgren2011explicit} gives the explicit connection between the GMRF and the Matérn GF; we refer the reader to \citet{lindgren2011explicit}, which provides the theoretical foundation for the proposed approximation. To simplify the presentation, we present in this section the resulting low-rank representation, and defer the technical details to Appendix~\ref{appendix:weak_approxiation}. 



Let us consider a partition of the domain $\Omega \subset \mathbb{R}^2$ into a set of non-overlapping triangles, i.e., $\Omega = \bigcup_i E_i$, where any two triangles share at most a common edge or vertex. 
The result of the discretisation is a mesh $\mathcal{G}_R=(\mathcal{V}_R,\mathcal{E}_R)$, with $R$ equal to the cardinality of the vertices  $\mathcal{V}_R = \{\bm r_1,\dots, \bm r_R\}$ and $\mathcal{E}_R$ being the set of the edges of the triangles. 
Based on the mesh $\mathcal{G}_R$, we define a set of basis functions $\bm \psi_R(\bm{s}) = (\psi_1(\bm{s}), \dots, \psi_R(\bm{s}))'$, where each $\psi_i: \mathbb{R}^2 \to \mathbb{R}$ is selected from the class of piecewise linear functions with compact support. These functions are constructed such that $\psi_i(\bm{r}_k) = 1$ if $i = k$, and $\psi_i(\bm{r}_k) = 0$ if $i \neq k$, for all $i, k = 1, \dots, R$. Additionally, we impose the condition that all basis functions have zero normal derivatives on the boundary of $\Omega$, which ensures that the SPDE approximation of the process $\eta_i(\bm{s}, t)$ exhibits the desired asymptotic properties, discussed in the sequel. A similar approach is presented in \cite{lindgren2011explicit}, to which we refer for further technical details. 

Let $\bm{\eta}_j(\mathcal{V}_R, t)$ denote the vector:
$$
\bm{\eta}_j(\mathcal{V}_R, t) = \left( \eta_j(\bm{r}_1, t), \dots, \eta_j(\bm{r}_R, t) \right)',
$$
for $j = 1, \dots, q$, where $\mathcal{V}_R = \{\bm{r}_1, \dots, \bm{r}_R\}$ is a finite set of spatial locations. 
Then, we define the full vector $\bm{\eta}(\mathcal{V}_R, t) \in \mathbb{R}^{qR}$ as:
$$
\bm{\eta}(\mathcal{V}_R, t) = \left( \bm{\eta}_1(\mathcal{V}_R, t)', \dots, \bm{\eta}_q(\mathcal{V}_R, t)' \right)'.
$$
The extension to the case where each component has a different mesh, i.e. $\mathcal{G}_R^j = (\mathcal{V}_R^j, \mathcal{E}_R^j)$ for $j = 1, \dots, q$, is straightforward. Here, for simplicity and without loss of generality, we assume that the latent process components share the same finite element mesh $\mathcal{G}_R$. In Section~\ref{sec:case_study}, we provide an example with disjoint meshes. Similarly, let $\bm z_j(\mathcal{V}_R, t)$ for $j = 1,\dots, q$ be the vector $\bm z_j(\mathcal{V}_R, t) = (z_{j}(\bm r_1,t),\dots, z_{j}(\bm r_R,t))'$ and $\bm z(\mathcal{V}_R, t)$ be the vector $\bm z(\mathcal{V}_R, t) = (\bm z_1(\mathcal{V}_R, t)', \dots, \bm z_q(\mathcal{V}_R, t)')'$. 

Let $\mathcal{H}^1(\Omega)$ denote the first-order Sobolev space over the spatial domain $\Omega$. Consider the finite-dimensional subspace $\mathcal{H}_R^1(\Omega) \subset \mathcal{H}^1(\Omega)$ spanned by basis functions $\{\psi_i\}_{i=1}^R$. We approximate each component of the innovation term in \eqref{eq:state_space2} using the Galerkin solution of the SPDE defining the Matérn GF (see the Appendix \ref{appendix:weak_approxiation}, Equation \eqref{eq:partial_differential_equation}), projected onto $\mathcal{H}_R^1(\Omega)$. This yields the following approximation:
\begin{equation}\label{eq:approximation_state_space2}
 \bm z(\bm s, t) \approx \bm F \bm z(\bm s, t-1) + \bm \Psi_R(\bm s)\bm \eta(\mathcal{V}_R, t),
\end{equation}
where $ \bm \Psi_R(\bm s) = \bm I_p \otimes \bm \psi_R(\bm s)'$,  $ \bm \psi_R(\bm s) = (\psi_1(\bm s),\dots,  \psi_R(\bm s))'$ and $\otimes$ denotes the Kronecker product.

We propose to leverage \eqref{eq:approximation_state_space2} to obtain a low-rank approximation of model \eqref{eq:state_space1}-\eqref{eq:state_space2}. That is, considering  \eqref{eq:approximation_state_space2} holding as equality, we replace \eqref{eq:state_space1}-\eqref{eq:state_space2} with the following LR-SSM:
\begin{align}
           & \bm{y}(\bm{s},t) = \bm{X}'(\bm{s},t)\bm{\beta} + \bm W\bm \Psi_R(\bm s)\bm z(\mathcal{V}_R,t)'+ \bm{\varepsilon}(\bm{s},t) \label{eq:proposed_ssm1}\\
           & \bm{z}(\mathcal{V}_R, t) = (\bm F\otimes \bm I_R) \bm{z}(\mathcal{V}_R, t - 1) + \bm{\eta}(\mathcal{V}_R, t). \label{eq:proposed_ssm2}
\end{align}
In Equation \eqref{eq:proposed_ssm2}, the continuous latent field $\bm{z}(\bm{s}, t)$ from \eqref{eq:state_space1} is replaced by a low-rank approximation of the form $\bm{\Psi}_R(\bm{s}) \bm{z}(\mathcal{V}_R, t)'$, where the reduction ratio is defined as $LR = \frac{|\mathcal{V}_R|}{m} \times 100\%$. Comparing \eqref{eq:state_space1} with \eqref{eq:proposed_ssm1}, the approximation introduces a model-based error. In Section~\ref{sec:theoretical_results}, we show that the model-based error can be controlled by the number of vertices $R$. As expected, a finer finite element mesh provides increased inference accuracy.

We summarise here the advantages of using the model defined in equations \eqref{eq:proposed_ssm1}--\eqref{eq:proposed_ssm2} to make inference on model \eqref{eq:state_space1}--\eqref{eq:state_space2}. 
First, the low-rank approximation reduces the computational complexity of the Kalman filter from $\mathcal{O}(m^3T)$ to $\mathcal{O}(mR^2T)$, gaining efficiency for $R \ll m$. The practitioner can therefore control the balance between computational cost and accuracy by choosing the value of $R$. Second, it preserves the first- and second-order moments of the original process in the limit, thereby enabling scalable inference and prediction in high-dimensional spatio-temporal settings. Third, once the random coefficients $\bm z (\mathcal{V}_R, t)$ are estimated, evaluating the latent process at new locations has a linear computational cost. This enables efficient construction of spatial maps of the observed processes, making the proposed model competitive even when $R=m$ ($LR = 100\%$). Details on model estimation and spatial mapping are provided in Section~\ref{sec:estimation}. Most model parameters can be estimated via closed-form expressions. For the rescaling parameters, $k_i$ for $i = 1,\dots,q$, although no closed-form solution is available, the assumption of independent innovation components allows for efficient minimisation whose complexity grows linearly with the dimension of the latent component $q$. This makes the estimation procedure particularly efficient 
as compared to other SSMs such as those in \cite{calculli2015maximum}.

\subsection{Identifiability, Convergence, and Approximation Bounds}
\label{sec:theoretical_results}
We begin by analysing the theoretical properties of the proposed approximation. Since the true data-generating process is unknown, we treat the LR-SSM~\eqref{eq:proposed_ssm1}–\eqref{eq:proposed_ssm2} as the underlying model. In this way, we first establish point identification in a distributional sense, followed by a weak convergence result for the low-rank approximation. Finally, we quantify the approximation error by deriving a uniform bound on the difference between the full-rank SSM and its low-rank counterpart.

\begin{assumption}
\label{ass:full_rank}
For a subset $\mathcal{S} \subset \Omega$, the limit $ \lim_{T \to \infty} \frac{1}{T} \int_{\mathcal{S}} \bm{X}(\bm{s}, t)\, \text{d}\bm{s} = L_{\bm{X}}$ exists and is finite, and the matrix $L_{\bm{X}} L_{\bm{X}}'$ has full rank.
\end{assumption}
Assumption~\ref{ass:full_rank} imposes a standard full-rank condition on the fixed-effects component of the model, ensuring identifiability of the regression coefficient vector $\bm\beta$.
\begin{assumption}\label{ass:identifiability}
     $(i)$ $f_1 > f_2 > \dots > f_q$; $(ii)$ $\bm W$ has full column rank, with each column scaled so that its first non-zero element is positive; $(iii)$ $\bm\Psi_R(\bm s)=\bm I_p\otimes\bm\psi_R(\bm s)'$ is known and has full row rank. 
\end{assumption}
Assumption~\ref{ass:identifiability} imposes restrictions on the parameter space, excluding alternative representations of the LR-SSM that may result from permuting the order or changing the sign of latent components. The following identifiability result is established, with the proof provided in Appendix~\ref{appendix:identification}. We consider point identification in a distributional sense, where model parameters are uniquely determined by the joint distribution of the observables \citep{lewbel2019identification}.

\begin{theorem}\label{thm:identifiability}
Let Assumptions~\ref{ass:full_rank}--\ref{ass:identifiability}, hold, and consider the data generated according to the LR-SSM~\eqref{eq:proposed_ssm1}–\eqref{eq:proposed_ssm2}, with parameter set $\Pi$. Assume that the process $\{\bm y(\bm s, t) : (\bm s, t) \in \mathcal S \times \mathbb N\}$ is observed over a spatially dense set $\mathcal S \subset \Omega$ and over an arbitrarily long time horizon. Then, the parameter set $\Pi$ is identifiable, meaning that no two distinct parameter sets $\Pi$ and $\Pi^{\star}$ can generate the same distribution of the observed process. 
\end{theorem}


The SSM models described in \eqref{eq:state_space1}–\eqref{eq:state_space2} provide a continuous representation of the observed random process, capturing its inherent continuity. We now analyse the estimation errors that arise when the LR-SSM is employed to infer data generated by the true SSM.

First, we discuss the convergence of the low-rank approximation of the latent component. Let $x^R$ be a sequence of $L^2(\Omega)$-bounded Gaussian fields. As defined in \cite{lindgren2011explicit}, we say $x^R \xrightarrow{D\{L^2(\Omega)\}} x$ if for all $f, g \in L^2(\Omega)$,
\begin{align*}
\mathbb{E}[\langle f, x^R \rangle] \to \mathbb{E}[\langle f, x \rangle], \quad \text{and}\quad 
\operatorname{Cov}(\langle f, x^R \rangle, \langle g, x^R \rangle) \to \operatorname{Cov}(\langle f, x \rangle, \langle g, x \rangle).
\end{align*}

As $\mathcal{H}_R^1(\Omega)$ is dense in $\mathcal{H}^1(\Omega)$ as $R \to \infty$, Theorem 3 of \cite{lindgren2011explicit} ensures that for a fixed $t \in \{1, \dots, T\}$, it holds $\bm \Psi(\bm s)\bm \eta(\mathcal{V}_R,t) \xrightarrow{D\{L^2(\Omega^q)\}} \bm \eta(\bm s,t)$.  
Thanks to the Markovian structure of $\bm{z}(\bm{s}, t)$, over a finite time horizon $T$, this result implies that the approximate latent field converges to the true latent field as well, as stated in the following result.


\begin{theorem}[Weak Convergence of the Approximated State Process]\label{theo:z_convergeces}
For every $t = 1,\dots, T$ and $i=1,\dots, q$, let \(\eta_i(\bm s, t)\) satisfy the Matérn SPDE \eqref{eq:partial_differential_equation} with homogenous Neumann condition, and define $z_i(\bm s,t) = \sum_{r=0}^tf_i^{t-r}\eta_i(\bm s, r)$. Let \(\bm \eta_i(\mathcal{V}_R, t)\) be the vector of coefficients of the Galerkin solution of the SPDE in \(\mathcal H^1_{R}(\Omega,\kappa)\). Then
\[
\bm \Psi_R(\bm s) \bm z(\mathcal{V}_R, t) \xrightarrow{D\{L^2(\Omega^{qT})\}}\bm z(\bm s, t),
\]
where $\bm z(\mathcal{V}_R, t)$ are defined in terms of $\{\bm \eta_i(\bm s, t)\}_{i=1}^q$ in \eqref{eq:proposed_ssm2}. In particular,
\begin{align*}
\mathbb{E}[\bm \Psi_R(\bm s)\bm z(\mathcal{V}_R, t)] &\to \mathbb{E}[\bm z(\bm s, t)], \\
\operatorname{Cov}(\bm \Psi_R(\bm s)\bm z(\mathcal{V}_R, t), \bm \Psi_R(\tilde{\bm s})\bm z(\mathcal{V}_R, \tilde{t})) &\to \operatorname{Cov}(\bm z(\bm s, t), \bm z(\tilde{\bm s}, \tilde{t})).
\end{align*}
\end{theorem}
The proof is provided in Appendix~\ref{appendix:proof_z_convergence}, and relies on standard assumptions about the finite element mesh topology, namely that the minimal mesh angles are bounded away from zero and that the maximum edge length tends to zero. Importantly, the proposed model preserves the first- and second-order moments of the original process in the limit, thereby enabling scalable inference and prediction in high-dimensional spatio-temporal settings.

As a second result, we provide a bound on the observed process when the original model \eqref{eq:state_space1}–\eqref{eq:state_space2} is approximated by the model \eqref{eq:proposed_ssm1}–\eqref{eq:proposed_ssm2}.

\begin{theorem}[Bound on the Observable Process]\label{theo:approximation_yt}For every $t = 1,\dots, T$ and $\bm s\in \Omega$, let $\bm y(\bm s,t)$ be a random process satisfying the model \eqref{eq:state_space1}-\eqref{eq:state_space2} with set of parameter $\Pi$. Let  $\bm y^R(\bm s, t)$ the discretised version of $\bm y(\bm s,t)$ obtained by projecting each component of $\bm \eta_t$ into $\mathcal{H}_R(\Omega)$, so that $\{\bm y_t^R\}_{i=1}^T$ satisfies model \eqref{eq:proposed_ssm1}-\eqref{eq:proposed_ssm2} with the same parameter set $\Pi$. The following bound holds:
    \[
    \mathbb{E}\|\bm y_t - \bm y_t^R\|^2_2 \leq C \, \frac{1-\varphi^{2(t+1)}}{1-\varphi^2}\,  h^2,
    \]
    where $C$ is a constant that depends on $\|\bm{W}\|$ and $q$, $\varphi = \max_{i}|f_i| < 1$, $h$ denotes the maximum edge length of the finite element mesh used in the approximation, and $\|\cdot\|_{2}$ stands for the $L^2(\Omega)-$norm.
\end{theorem}
We prove this result in Appendix~\ref{appendix:proof_theo_approximation_yt}.  
The result in Theorem \ref{theo:parameter_bounds} shows that the error when LR-SSM and SSM are parametrised by the same parameter set $\Pi$, is uniformly bounded in time since \(\frac{1-\varphi^{2(t+1)}}{1-\varphi^{2}} \le \frac{1}{1-\varphi^{2}}\) as \(t \to \infty\). Thus, the discrepancy between the true and reduced–order outputs can never grow without bound; at worst, it is proportional to the finite element mesh width \(h\). The linear dependence on \(h\) reflects the usual accuracy–cost trade-off: a finer mesh reduces the error while demanding greater computational effort. The constant \(C\) depends on \(\|\bm{W}\|\) and \(q\), since the matrix \(\bm{W}\) modulates how the approximation error introduced by replacing \(\bm{z}\) with \(\bm{z}^R\) propagates to the observable output \(\bm{y}\).


Lastly, we obtain the following result concerning the parameter bias.

\begin{theorem}\label{theo:parameter_bounds}
    Under Assumptions~\ref{ass:full_rank}--\ref{ass:identifiability}, for a fixed $R$, the bias in the estimation of the vector parameters $\bm \pi = (\bm \beta', \bm \sigma',\operatorname{vec}(\bm W)', \bm f', k_1,\dots, k_q)'$ using the LR-SSM~\eqref{eq:proposed_ssm1}--\eqref{eq:proposed_ssm2} for data generated by the SSM~\eqref{eq:state_space1}--\eqref{eq:state_space2} is of order $\mathcal{O}(h)$, where $h$ denotes the maximum edge length of the finite element mesh used in the approximation. Specifically, 
    $$\bigl\|\bm \pi_R^{\star}-\bm \pi\bigr\| = \mathcal O(h)$$
where $\bm \pi_R^{\star}$ is the pseudo-true parameter vector minimising the Kullback-Leibler divergence ($D_{KL}$) between the true marginal law $\mathcal P_{\bm \pi}$ of $\bm y(\bm s, t)$ with parameter vector $\bm \pi$ and the misspecified LR-SSM marginal law $\mathcal P^{(R)}_{\bm \vartheta}$ of $\bm y^R(\bm s, t)$ with vector of parameter $\bm\vartheta$, i.e. $\bm \pi^\star_R = \operatorname{argmin}_{\bm\vartheta} D_{KL}(\mathcal P^{(R)}_{\bm \vartheta}||\mathcal P_{\bm \pi})$. Furthermore, the parameter $\bm \beta$ has no bias.
\end{theorem}
The proof is reported in Appendix~\ref{appendix:parameter_bound}. Together, Theorems~\ref{theo:z_convergeces}–\ref{theo:parameter_bounds} provide a comprehensive theoretical foundation supporting the use of the LR-SSM as a principled and efficient approximation to the full SSM, justifying its adoption in practical spatio-temporal inference and prediction tasks.


\section{Parameter estimation}\label{sec:estimation}
We estimate the parameter set $\Pi$ using a maximum likelihood (ML) approach, implemented via the EM algorithm. This method is particularly well-suited for models with latent structures. 
Our objective is to estimate $\Pi$ based on the observed data. 
Specifically, we observe each component $y_j(\bm{s}, t)$ at spatial locations $\mathcal{S}_{j,t} = \{\bm{s}_{i,t}\}_{i=1}^{m_{j,t}}$.

We assume that the mesh $\mathcal{G}_R$ 
—which defines the spatial domain for the latent process within the state-space model described in Equations~\eqref{eq:proposed_ssm1}–\eqref{eq:proposed_ssm2}—is given. The procedure used to construct this mesh is detailed in Section~\ref{sec:triangulation}.

The complete EM algorithmic steps, the closed-form expressions for the parameter updates, as well as the matrix formulation of the model, are presented in Appendix~\ref{appendix:EM}. We denote by $\bm Q_{\kappa_i}^{-1}$ the covariance matrix of $\bm \eta_i(\mathcal{V}_R,t)$, as specified in \eqref{eq:matern}. In Section \ref{sec:boundary}, we show how the precision matrix  $\bm Q_{\kappa_i}$ is constructed at linear cost using the FEM \citep{lindgren2011explicit}. This makes the maximisation procedure for the estimation of the parameter $\kappa_i$ with a linear cost instead of a cubic cost.  


Note that given the MLE of the parameter set $\Pi$, the plug-in predictions at new site $\bm s_0 \in \Omega$ and time $t = 1, \dots, T$ are given by
\begin{align}
           & \hat{\bm{y}}(\bm{s_0},t) = \bm{X}'(\bm{s_0},t)\hat{\bm{\beta}} + \hat{\bm W} \bm \Psi_R(\bm s_0) \hat{\bm z}^T(\mathcal{V}_R,t) \label{eq:prediction_mean_lssm}\\           
           & \bm \Sigma_{\hat{\bm{y}}}(\bm{s_0},t) = \hat{\bm W} \bm \Psi_R(\bm s_0) \operatorname{Var}[\hat{\bm z}^T(\mathcal{V}_R,t)] \bm \Psi_R(\bm s_0)' \hat{\bm W}' \label{eq:prediction_std_lssm}
\end{align}

\noindent where $\hat{\bm z}^T(\mathcal{V}_R, t)$ denotes the smoothed state process $\bm z(\mathcal{V}_R, t)$, with associated variance $\operatorname{Var}[\hat{\bm z}^T(\mathcal{V}_R, t)]$, estimated by the Kalman smoother variance $\bm P_t^T$ (see the Kalman smoother derivation in Appendix \ref{appendix:EM}). The vectors $\hat{\bm{\beta}}$ and $\hat{\bm W}$ are the MLE of $\bm \beta$ and $\bm W$, respectively. 
Equation \ref{eq:prediction_mean_lssm} results in the BLUP procedure, and it relies only on linear algebra operations with a linear computational cost of $\mathcal{O}(R)$.  

\subsection{Choice of the Mesh}\label{sec:triangulation}

Theorem~\ref{theo:z_convergeces} establishes the weak convergence $\bm \Psi_R(\bm s)\bm z(\mathcal{V}_R, t) \to \bm z(\bm s, t)$. This result holds under standard assumptions on the finite element mesh, say $\mathcal{G}_R = (\mathcal{V}_R, \mathcal{E}_R)$, specifically that the minimal mesh angles are bounded away from zero and the maximum edge length tends to zero.

Since the choice of triangulation significantly affects both the quality of the approximation and the asymptotic convergence, it is essential to construct finite element meshes that satisfy these topological criteria. In this section, we introduce a method for generating such meshes, involving $R$ vertices, where $R$ is chosen to balance computational tractability with accurate and reliable parameter estimation.

Let $\mathcal{S}$ denote the set of all observed locations, that is, $\mathcal{S} = \bigcup_{i=1}^p \bigcup_{t=1}^T\mathcal{S}_{i,t}$, where $\mathcal{S}_{i,t}$ is the set of locations at which variable $i$ is observed at time $t$. 

Among available options, Delaunay triangulations are commonly used to construct finite element meshes because they maximise the smallest angle among all possible triangulations of a given set of points, thereby avoiding the creation of overly narrow triangles. We construct an initial triangulation $\mathcal{G}_m$ by applying the Delaunay algorithm to the full set $\mathcal{S}$, where $m = |\mathcal{S}|$. To reduce the mesh to a fixed number $R$ of vertices, we apply an iterative procedure: at each step, the node associated with the smallest triangle area is removed, followed by local re-triangulation to preserve the Delaunay structure. This process is repeated until exactly $R$ vertices remain. The resulting triangulation defines the mesh $\mathcal{G}_R = (\mathcal{V}_R, \mathcal{E}_R)$. To ensure that the interior angles in the mesh are suitably bounded away from zero, we apply standard Laplacian smoothing \citep{field1988laplacian} to the triangular finite mesh $\mathcal{G}_R$. Laplacian smoothing is an effective ad hoc method for improving mesh quality, particularly in regions with varying element density, by adjusting vertex positions to produce a more regular triangulation.
The Laplacian smoothing algorithm for Delaunay triangulations is presented in pseudocode in Algorithm~\ref{alg:laplacian_smoothing}. It produces a Laplacian-smoothed version of the Delaunay triangulation, denoted by $\mathcal{G}_R^* = (\mathcal{V}_R^*, \mathcal{E}_R^*)$. The resulting set of smoothed vertices $\mathcal{V}_R^*$ can then be used in place of $\mathcal{V}_R$ in the state equation of the SSM model~\eqref{eq:proposed_ssm1}–\eqref{eq:proposed_ssm2}. For additional details and visual illustrations demonstrating the effectiveness of this approach in improving angle quality relative to standard Delaunay triangulations, we refer the reader to \citet{field1988laplacian}.

\begin{algorithm}[H]
\caption{Laplacian smoothing algorithm for Delaunay triangulations. \label{alg:laplacian_smoothing}}
\KwIn{Triangulation $\mathcal{G}_R = (\mathcal{V}_R, \mathcal{E}_R)$; angle threshold $\theta_{\min}$}
\KwOut{Triangulation with all angles $\geq \theta_{\min}$}
\While{any triangle angle $< \theta_{\min}$}{
  \ForEach{interior vertex $\bm{r} \in \mathcal{V}_R$}{
    Let $\{\bm{r}_1, \ldots, \bm{r}_k\}$ be the neighbours of $\bm{r}$\;
    Compute $\bm{r}^* \gets \frac{1}{k} \sum_{i=1}^{k} \bm{r}_i$\;
    \eIf{$\bm{r}^*$ preserves Delaunay and connectivity}{
      Update position: $\bm{r} \gets \bm{r}^*$\;
    }{
      Remove $\bm{r}$; locally retriangulate\;
      Insert vertex at $\bm{r}^*$\;
    }
  }
}
\end{algorithm}

\subsection{Approximation of the Precision Matrix}\label{sec:boundary}

In this section, we briefly recall how the approximate precision matrix $\bm{Q}$ is obtained within the SPDE framework of \citet{lindgren2011explicit}; further details are given in
Appendix~\ref{appendix:weak_approxiation}. 

Let $\mathcal{G}_R =(\mathcal{V}_R,\mathcal{E}_R)$ be a triangulation of the spatial domain $\Omega$ with $R = |\mathcal{V}_R|$ vertices associated with piecewise–linear basis functions $\{\psi_i\}_{i=1}^{R}$ of the finite element space. Let $\bm{C}$ and $\bm{G}$ denote the mass matrix and stiffness matrix, respectively, defined as
\[
\bm{C}_{ij}= \langle\psi_i,\psi_j\rangle,
\qquad
\bm{G}_{ij}= \langle\nabla\psi_i,\nabla\psi_j\rangle.
\]
We then define the matrix
\[
\bm{K}_{\kappa_i^2}= \kappa_i^{2}\bm{C}+\bm{G}.
\]

Using the Neumann boundary conditions, the finite-dimensional representations of the solutions to Equation \ref{eq:spde_full} have precision

\begin{equation}\label{eq:precision_gmrf}
\bm{Q}_{k_i}
= \bm{K}_{\kappa_i^2}\,\bm{C}^{-1}\,\bm{K}_{\kappa^2}.  
\end{equation}

The matrix $\bm{C}^{-1}$ is dense, which leads to a dense precision matrix in the resulting model. To address this issue, \citet[Appendix C.5]{lindgren2011explicit} show that $\bm{C}$ can be approximated by a diagonal matrix $\widetilde{\bm{C}}$, where $\widetilde{C}_{ii} = \langle \psi_i, 1 \rangle \,$. This approximation yields a sparse precision matrix, resulting in a GMRF representation of the original Matérn GF with rescale $k_i$. Because each basis function overlaps only a fixed, small number of neighbours, $\widetilde{\bm{C}}$, $\bm{G}$, $\bm{K}_{\kappa_i^2}$ contain $\mathcal{O}(R)$ non-zero entries. Thus, all matrix products can be assembled with linear complexity $O(R)$. Due to the independence assumption of the $q$ components of the random vector $\bm{\eta}(\mathcal{V}_R, t)$, the resulting precision matrix is $\bm{Q} = \operatorname{blockdiag}(\bm{Q}_{k_1}, \dots, \bm{Q}_{k_q})$

The result in \eqref{eq:precision_gmrf} is derived using Neumann boundary conditions to solve the SPDE \eqref{eq:spde_full}. We summarise the main considerations about the boundary effect and the boundary-mitigation strategy in the next paragraph and refer the reader to Appendix~\ref{appendix:boundary} for further details. 

\subsubsection{Boundary effects}
Imposing Neumann boundary conditions on the field $\boldsymbol{\eta}(\bm{s},t)$ induces a folding effect in the corresponding covariance function. This behaviour was analysed in detail by \citet{lindgren2011explicit}; Appendix~\ref{appendix:boundary} revisits the main arguments and demonstrates that, at distances greater than twice the Matérn range parameter from the domain boundary, the folded covariance is practically indistinguishable from the stationary Matérn covariance. We leverage this result by enlarging the vertex set through the addition of auxiliary boundary vertices; that is, we work on the augmented finite element mesh $\mathcal{V}_R^{\ast} = \mathcal{V}_R \cup \mathcal{V}_R^{\mathrm{aux}}$ where $\mathcal{V}_R^{\mathrm{aux}}$ is chosen so that every vertex in $\mathcal{V}_R$ lies at least two Matérn ranges away from the domain edge. On this extended finite element mesh, we build the SPDE-based precision matrix $\bm{Q}_{k_i}^{\ast}$ as in \ref{eq:precision_gmrf}. Finally, the target precision matrix is recovered as the principal submatrix
\[
\bm{Q}_{k_i}= \bm{Q}_{k_i}^{\ast}[\,\mathcal{I},\mathcal{I}\,],
\]
where $\mathcal{I}$ indexes the original vertices $\mathcal{V}_R$.

\section{Simulations} \label{sec:simulation}

In this section, we evaluate the finite-sample performance of the proposed LR-SSM. To this end, we simulate model~\eqref{eq:state_space1}–\eqref{eq:state_space2} using the parameter vector denoted by $\Pi^0$. The inference is carried out using the proposed LR-SSM~\eqref{eq:proposed_ssm1}–\eqref{eq:proposed_ssm2}. 
Indeed, Theorem \ref{theo:approximation_yt} shows that the approximation error when using model \eqref{eq:proposed_ssm1}-\eqref{eq:proposed_ssm2} for inference on a data-generating process of model~\eqref{eq:state_space1}-\eqref{eq:state_space2} remains uniformly bounded in time. We consider a realistic setup, where the observed process is of dimension $p=3$ and a latent space of dimension $q=2$, allowing for a fully heterotopic case. 

\subsection{Experimental Design}

In order to study the identifiability of the model defined by \eqref{eq:proposed_ssm1}-\eqref{eq:proposed_ssm2}, parametrised by $\Pi^0$, synthetic data are generated using the SSM defined by equations~\eqref{eq:state_space1}–\eqref{eq:state_space2}. Let $\mathcal{L}_\delta = \left\{ (i\delta, j\delta) \in \mathbb{R}^2 : i, j \in \{0, 1, \dots, 24\} \right\}$ denote a $25 \times 25$ regular lattice on the domain $\Omega = [0,1]^2$, with spacing $\delta = 1/25$ between adjacent points. For each variable $i \in \{1, \dots, p\}$, the training set $\mathcal{S}_i \subset \mathcal{L}_\delta$ is constructed by selecting a random subset of $m$ locations. The remaining points, $\mathcal{S}_i^* = \mathcal{L}_\delta \setminus \mathcal{S}_i$, are used to evaluate the model’s predictive performance. 

\begin{figure}
    \centering
    \includegraphics[scale=0.45]{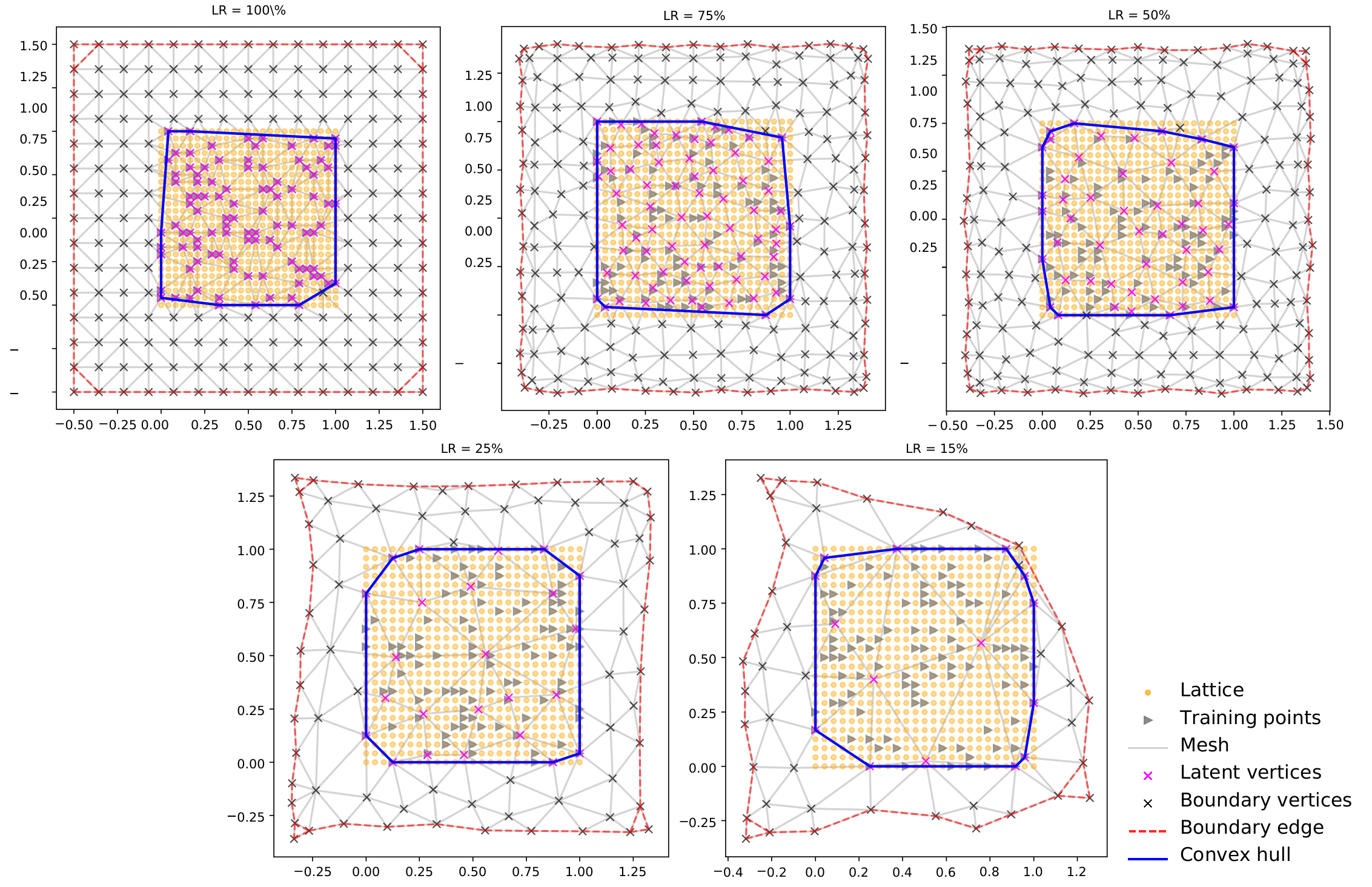}
    \caption{Simulated $m = 100$ spatial points location on the domain \(\Omega = [0,1]^2\), along with the resulting finite element mesh for $LR= 100\%, 75\%, 50\%, 25\%$ and $15\%$(from left to right, top to bottom). Laplacian smoothing is performed providing a refined vertices (magenta crosses) of the original Delaunay triangulation of the observed points (light grey triangles). The predictive performances are computed on a regular $25 \times 25$ lattice $\mathcal{L}_\delta$ (light orange points). The observed points are shown in light grey. The boundary vertices are shown in grey. }
    \label{fig:lowrank_mesh}
\end{figure}

We consider three different cases, corresponding to \(m = 100\), 225, and 400 spatial locations while  $LR$ is set to 100\%, 75\%, 50\%, 25\% and 15\%. As an example, Figure \ref{fig:lowrank_mesh} shows the generated finite element mesh for the simulations when $m = 100$. Except for the case $LR = 100\%$, all the other cases consider the Laplacian smoothing of the Delaunay triangulation. Details on Laplacian smoothing and boundary effects are in Section \ref{sec:triangulation} and Section \ref{sec:boundary}. Moreover, we consider the effect of time series length by considering \(T = 50\) and \(100\). 

Summing up, for each rank $R$, the simulation campaign is based on six different simulation setups. For each simulation setup, $M =100$ Monte Carlo replications are performed.

The data-generating process is specified by $\Pi^0 = \{ \bm \beta^0, \sigma^{2,0}, \bm f^0, \bm W^0, k_1^0, \dots, k_q^0\}$ with the following parameter values: \(\bm\beta^0 = (1, 2, -1)'\), and
\[
\bm X'(\bm s, t) =
\begin{bmatrix}
    x_1(\bm s, t) & 0 & 0 \\
    0 & x_2(\bm s, t) & 0 \\
    0 & 0 & x_3(\bm s, t) \\
\end{bmatrix},
\]
where each \(x_i(\bm s, t)\) is independently drawn from a standard normal distribution \(\mathcal{N}(0,1)\), for all spatial locations \(\bm s \in \mathcal{L}_\delta\), time steps \(t = 1, \dots, T\), and for \(i = 1, 2, 3\).
The variance parameters are set to \(\bm\sigma^{2,0} = (0.5, 1.5, 1)'\). The autoregressive coefficients are defined as \(\bm f = (0.85, -0.5)'\), and the $p \times q$ interaction matrix is given by
\[
\bm W^0 =
\begin{bmatrix}
    0.5 & 1 \\
    0.5 & 0.25 \\
    0.2 & 0.8 \\
\end{bmatrix}.
\]
The spatial range parameters are set to \(\bm\kappa^0 = (\kappa_1^0, \kappa_2^0)' = (7\sqrt{8}, 2\sqrt{8})'\). 
The latent variables \(\bm z(\bm s, t)\) are simulated on the same spatial grid with initial state drawn from \(\bm z(\bm s, 0) \sim \mathcal{N}(1, 1)\).

\subsection{EM estimation}
Once a single dataset $j$, for $j = 1, \dots, M$ is simulated, the EM algorithm is executed with particular starting values. Specifically, as starting values for the estimation procedure, we use a random point uniformly drawn from \([0,1]^2\) for \(\bm \beta\), a random point in \((0.2, 0.8) \times (-0.2, -0.8)\) for \(\bm f\), each element of \(\bm W\) is sampled independently from \(\mathcal{U}[0.2, 2]\), each \(\kappa_i\) is drawn from \(\mathcal{U}(\sqrt{8}, 7\sqrt{8})\), and each \(\sigma^2_i\) from \(\mathcal{U}[0.1, 2]\). 

The maximum iteration number of the EM algorithm is set to 300, while the exit condition is based on the convergence criteria $| \text{logL}_{k+1} - \text{logL}_k |/|\text{logL}_k| < 0.0001$ where $\text{logL}_k$ denote the log-likelihood at iteration $k$. 

\subsection{Simulation results}
We evaluate the models by comparing their parameter estimates using the bias and root mean square error (RMSE) where for each $\bm \theta \in \Pi$ the RMSE is computed as $\sqrt{N^{-1}\sum_{i=1}^M \|\hat{\bm\theta}^i-\bm \theta_0\|^2}$ where $\hat{\bm\theta^i}$ corresponds to the estimation of the parameter $\bm \theta$ in the $i$th Monte Carlo simulation, $\|\cdot\|$ is the euclidean norm and $\bm \theta_0$ is the true parameter value.  In addition, we assess predictive performance on the validation set $\mathcal{S}_i^*$ for $i = 1, \dots, p$, by comparing the observed process $y_i(\bm{s}, t)$ with its estimate $\hat{y}_i(\bm{s}, t)$. Prediction errors are defined as $e_i(\bm{s}, t) = y_i(\bm{s}, t) - \hat{y}_i(\bm{s}, t)$, for $\bm{s} \in \mathcal{S}_i^*$. The different LR-SSM versions are then compared in terms of RMSE computed over these validation errors.  

Tables~\ref{tab:consolidated_results_R100}, \ref{tab:consolidated_results_R50} presents the bias and RMSE of the estimated parameters under each combination of \(m\) and \(T\) for $LR = 100\%, 75\%$ and $LR = 15\%$ respectively (the simulation results for the cases with $LR = 75\%$ and $LR = 25\%$ are showed in the Supplementary material). The results indicate that estimation accuracy improves as the number of spatial locations and time steps increases, as evidenced by decreasing RMSE and absolute bias values across all parameters. The ML estimator of \(\bm\beta\) has an RMSE close to zero, improving with increasing observed data both in \(T\) and $m$. The noise variances \(\bm\sigma^2\) show more pronounced improvement with increasing \(m\). Spatial range parameters \(\bm\kappa\) are more challenging to estimate accurately; although RMSE decreases with finer grids, a noticeable bias persists. This is a known issue in spatial statistics, as range parameters often exhibit weak identifiability, particularly when data are limited in spatial resolution or extent \citep{zhang2004inconsistent}. Moreover, the RMSE do not decrease fast to zero and maybe do not converge for $T \to \infty$. Note that this is not surprising as it is consistent with asymptotic theory, which requires consistent initial estimates. The ML estimator of \(\bm W\) has small RMSE, and the accuracy improves with increasing time length $T$. Similar behaviour is shown by the estimator of the temporal autocorrelation parameters \(\bm f\). 

\begin{table}[H]
\centering
\resizebox{\textwidth}{!}{%
\begin{tabular}{l rr rr rr rr rr rr}
\toprule
& \multicolumn{6}{c}{\textbf{T = 50}} & \multicolumn{6}{c}{\textbf{T = 100}} \\
\cmidrule(lr){2-7} \cmidrule(lr){8-13}
& \multicolumn{2}{c}{\textbf{m = 100}} & \multicolumn{2}{c}{\textbf{m = 225}} & \multicolumn{2}{c}{\textbf{m = 400}} & \multicolumn{2}{c}{\textbf{m = 100}} & \multicolumn{2}{c}{\textbf{m = 225}} & \multicolumn{2}{c}{\textbf{m = 400}} \\
\cmidrule(lr){2-3} \cmidrule(lr){4-5} \cmidrule(lr){6-7} \cmidrule(lr){8-9} \cmidrule(lr){10-11} \cmidrule(lr){12-13}
\textbf{Parameter} & \textbf{Bias} & \textbf{RMSE} & \textbf{Bias} & \textbf{RMSE} & \textbf{Bias} & \textbf{RMSE} & \textbf{Bias} & \textbf{RMSE} & \textbf{Bias} & \textbf{RMSE} & \textbf{Bias} & \textbf{RMSE} \\
\midrule
$\beta_1$ & 0.002 & \multirow{3}{*}{0.017} & 0.000 & \multirow{3}{*}{0.011} & 0.001 & \multirow{3}{*}{0.008} & 0.000 & \multirow{3}{*}{0.012} & 0.001 & \multirow{3}{*}{0.007} & -0.001 & \multirow{3}{*}{0.006} \\
$\beta_2$ & 0.001 & & 0.001 & & 0.001 & & 0.003 & & 0.001 & & 0.000 & \\
$\beta_3$ & -0.008 & & -0.001 & & -0.001 & & 0.001 & & 0.000 & & 0.000 & \\
\midrule
$\sigma_1^2$ & 0.160 & \multirow{3}{*}{0.193} & 0.067 & \multirow{3}{*}{0.105} & 0.013 & \multirow{3}{*}{0.048} & 0.150 & \multirow{3}{*}{0.180} & 0.060 & \multirow{3}{*}{0.096} & 0.014 & \multirow{3}{*}{0.046} \\
$\sigma_2^2$ & 0.278 & & 0.165 & & 0.079 & & 0.266 & & 0.153 & & 0.079 & \\
$\sigma_3^2$ & 0.121 & & 0.055 & & 0.019 & & 0.114 & & 0.052 & & 0.013 & \\
\midrule
$k_1$ & -3.526 & \multirow{2}{*}{0.207} & -3.842 & \multirow{2}{*}{0.200} & -4.767 & \multirow{2}{*}{0.262} & -3.097 & \multirow{2}{*}{0.174} & -3.084 & \multirow{2}{*}{0.159} & -4.386 & \multirow{2}{*}{0.236} \\
$k_2$ & 0.390 & & -0.368 & & -0.414 & & 0.619 & & -0.185 & & -0.355 & \\
\midrule
$W_1$ & 0.018 & \multirow{6}{*}{0.177} & -0.022 & \multirow{6}{*}{0.223} & 0.005 & \multirow{6}{*}{0.259} & 0.051 & \multirow{6}{*}{0.172} & -0.002 & \multirow{6}{*}{0.206} & 0.041 & \multirow{6}{*}{0.280} \\
$W_2$ & 0.143 & & 0.226 & & 0.191 & & 0.154 & & 0.215 & & 0.197 & \\
$W_3$ & -0.026 & & -0.032 & & -0.025 & & 0.014 & & -0.009 & & 0.022 & \\
$W_4$ & 0.007 & & 0.046 & & 0.046 & & 0.014 & & 0.043 & & 0.035 & \\
$W_5$ & 0.018 & & 0.004 & & 0.020 & & 0.036 & & 0.009 & & 0.030 & \\
$W_6$ & 0.121 & & 0.180 & & 0.149 & & 0.127 & & 0.177 & & 0.165 & \\
\midrule
$f_1$ & -0.067 & \multirow{2}{*}{0.133} & -0.052 & \multirow{2}{*}{0.094} & -0.083 & \multirow{2}{*}{0.169} & -0.082 & \multirow{2}{*}{0.121} & -0.056 & \multirow{2}{*}{0.077} & -0.074 & \multirow{2}{*}{0.141} \\
$f_2$ & -0.021 & & -0.007 & & -0.015 & & -0.008 & & -0.006 & & -0.017 & \\
\midrule
$\text{RMSE}_{\text{train}}$ & 1.022 &  & 0.989 &  & 0.964 &  & 1.015 &  & 0.983 &  & 0.961 &  \\
$\text{RMSE}_{\text{test}}$ & 1.233 &  & 1.178 &  & 1.143 &  & 1.217 &  & 1.158 &  & 1.124 &  \\
Runtime (hours) & 0.11 &  & 0.49 &  & 1.65 &  & 0.15 &  & 0.74 &  & 3.02 & \\
\bottomrule
\end{tabular}%
}
\caption{$LR = 100\%$. Bias and RMSE of the estimated parameters are reported for different numbers of time steps ($T$) and spatial locations ($m$). $RMSE_{\text{train}}$ and $RMSE_{\text{test}}$ denote the average prediction RMSEs computed on the training set $\mathcal{S}_i$ and the test set $\mathcal{S}_i^*$, respectively, across $M = 100$ Monte Carlo replications. Time indicates the average computation time over replications.
\label{tab:consolidated_results_R100}}
\end{table}

\begin{table}[H]
\centering
\resizebox{\textwidth}{!}{%
\begin{tabular}{l rr rr rr rr rr rr}
\toprule
& \multicolumn{6}{c}{\textbf{T = 50}} & \multicolumn{6}{c}{\textbf{T = 100}} \\
\cmidrule(lr){2-7} \cmidrule(lr){8-13}
& \multicolumn{2}{c}{\textbf{m = 100}} & \multicolumn{2}{c}{\textbf{m = 225}} & \multicolumn{2}{c}{\textbf{m = 400}} & \multicolumn{2}{c}{\textbf{m = 100}} & \multicolumn{2}{c}{\textbf{m = 225}} & \multicolumn{2}{c}{\textbf{m = 400}} \\
\cmidrule(lr){2-3} \cmidrule(lr){4-5} \cmidrule(lr){6-7} \cmidrule(lr){8-9} \cmidrule(lr){10-11} \cmidrule(lr){12-13}
\textbf{Parameter} & \textbf{Bias} & \textbf{RMSE} & \textbf{Bias} & \textbf{RMSE} & \textbf{Bias} & \textbf{RMSE} & \textbf{Bias} & \textbf{RMSE} & \textbf{Bias} & \textbf{RMSE} & \textbf{Bias} & \textbf{RMSE} \\
\midrule
$\beta_1$ & -0.002 & \multirow{3}{*}{0.018} & 0.000 & \multirow{3}{*}{0.011} & 0.001 & \multirow{3}{*}{0.008} & -0.004 & \multirow{3}{*}{0.013} & -0.001 & \multirow{3}{*}{0.007} & 0.000 & \multirow{3}{*}{0.006} \\
$\beta_2$ & -0.004 & & 0.002 & & -0.001 & & 0.004 & & -0.001 & & -0.001 & \\
$\beta_2$ & -0.004 & & 0.001 & & 0.001 & & -0.001 & & -0.001 & & 0.001 & \\
\midrule
$\sigma_1^2$ & 0.435 & \multirow{3}{*}{0.304} & 0.274 & \multirow{3}{*}{0.190} & 0.199 & \multirow{3}{*}{0.137} & 0.419 & \multirow{3}{*}{0.298} & 0.256 & \multirow{3}{*}{0.183} & 0.180 & \multirow{3}{*}{0.125} \\
$\sigma_2^2$ & 0.324 & & 0.207 & & 0.149 & & 0.332 & & 0.214 & & 0.141 & \\
$\sigma_3^2$ & 0.153 & & 0.083 & & 0.053 & & 0.142 & & 0.073 & & 0.043 & \\
\midrule
$k_1$ & -0.527 & \multirow{2}{*}{0.316} & -1.407 & \multirow{2}{*}{0.145} & -2.224 & \multirow{2}{*}{0.137} & -0.295 & \multirow{2}{*}{0.220} & -0.788 & \multirow{2}{*}{0.096} & -1.426 & \multirow{2}{*}{0.088} \\
$k_2$ & -0.228 & & -0.463 & & -0.396 & & 0.002 & & -0.412 & & -0.280 & \\
\midrule
$W_1$ & 0.029 & \multirow{6}{*}{0.196} & -0.018 & \multirow{6}{*}{0.216} & -0.016 & \multirow{6}{*}{0.203} & 0.081 & \multirow{6}{*}{0.195} & 0.003 & \multirow{6}{*}{0.220} & -0.001 & \multirow{6}{*}{0.218} \\
$W_2$ & 0.170 & & 0.214 & & 0.197 & & 0.153 & & 0.227 & & 0.222 & \\
$W_3$ & -0.006 & & -0.033 & & -0.025 & & 0.036 & & -0.006 & & -0.004 & \\
$W_4$ & 0.047 & & 0.067 & & 0.064 & & 0.043 & & 0.074 & & 0.066 & \\
$W_5$ & 0.014 & & -0.002 & & 0.000 & & 0.038 & & 0.005 & & 0.004 & \\
$W_6$ & 0.139 & & 0.177 & & 0.166 & & 0.136 & & 0.190 & & 0.188 & \\
\midrule
$f_1$ & -0.036 & \multirow{2}{*}{0.097} & -0.027 & \multirow{2}{*}{0.083} & -0.036 & \multirow{2}{*}{0.072} & -0.056 & \multirow{2}{*}{0.131} & -0.031 & \multirow{2}{*}{0.058} & -0.039 & \multirow{2}{*}{0.062} \\
$f_2$ & 0.002 & & -0.004 & & -0.013 & & -0.001 & & -0.012 & & -0.005 & \\
\midrule
$\text{RMSE}_{\text{train}}$ & 1.088 & & 1.042 &  & 1.022 & & 1.084 &  & 1.038 &  & 1.015 &  \\
$\text{RMSE}_{\text{test}}$ & 1.230 &  & 1.169 &  & 1.132 &  & 1.223 &  & 1.158 &  & 1.121 &  \\
Runtime (hours) & 0.05 &  & 0.17 & & 0.64 &  & 0.05 &  & 0.23 & & 0.88 & \\
\bottomrule
\end{tabular}%
}
\caption{\label{tab:consolidated_results_R50}$LR = 50\%$. Bias and RMSE of the estimated parameters are reported for different numbers of time steps ($T$) and spatial locations ($m$). $RMSE_{\text{train}}$ and $RMSE_{\text{test}}$ denote the average prediction RMSEs computed on the training set $\mathcal{S}_i$ and the test set $\mathcal{S}_i^*$, respectively, across $M = 100$ Monte Carlo replications. Time indicates the average computation time over replications.}
\end{table}

\begin{table}[H]
\centering
\resizebox{\textwidth}{!}{%
\begin{tabular}{l rr rr rr rr rr rr}
\toprule
& \multicolumn{6}{c}{\textbf{T = 50}} & \multicolumn{6}{c}{\textbf{T = 100}} \\
\cmidrule(lr){2-7} \cmidrule(lr){8-13}
& \multicolumn{2}{c}{\textbf{m = 100}} & \multicolumn{2}{c}{\textbf{m = 225}} & \multicolumn{2}{c}{\textbf{m = 400}} & \multicolumn{2}{c}{\textbf{m = 100}} & \multicolumn{2}{c}{\textbf{m = 225}} & \multicolumn{2}{c}{\textbf{m = 400}} \\
\cmidrule(lr){2-3} \cmidrule(lr){4-5} \cmidrule(lr){6-7} \cmidrule(lr){8-9} \cmidrule(lr){10-11} \cmidrule(lr){12-13}
\textbf{Parameter} & \textbf{Bias} & \textbf{RMSE} & \textbf{Bias} & \textbf{RMSE} & \textbf{Bias} & \textbf{RMSE} & \textbf{Bias} & \textbf{RMSE} & \textbf{Bias} & \textbf{RMSE} & \textbf{Bias} & \textbf{RMSE} \\
\midrule
$\beta_1$ & -0.001 & \multirow{3}{*}{0.018} & -0.002 & \multirow{3}{*}{0.011} & -0.001 & \multirow{3}{*}{0.008} & 0.000 & \multirow{3}{*}{0.012} & 0.000 & \multirow{3}{*}{0.008} & -0.001 & \multirow{3}{*}{0.006} \\
$\beta_2$ & -0.005 & & -0.001 & & 0.000 & & 0.002 & & 0.001 & & -0.001 & \\
$\beta_3$ & -0.001 & & 0.001 & & 0.000 & & 0.001 & & -0.002 & & 0.000 & \\
\midrule
$\sigma_1^2$ & 0.813 & \multirow{3}{*}{0.562} & 0.556 & \multirow{3}{*}{0.381} & 0.416 & \multirow{3}{*}{0.283} & 0.804 & \multirow{3}{*}{0.553} & 0.541 & \multirow{3}{*}{0.375} & 0.394 & \multirow{3}{*}{0.273} \\
$\sigma_2^2$ & 0.582 & & 0.403 & & 0.301 & & 0.577 & & 0.412 & & 0.303 & \\
$\sigma_3^2$ & 0.300 & & 0.179 & & 0.125 & & 0.289 & & 0.169 & & 0.113 & \\
\midrule
$k_1$ & -8.255 & \multirow{2}{*}{0.502} & -0.062 & \multirow{2}{*}{0.311} & 7.295 & \multirow{2}{*}{0.572} & -6.818 & \multirow{2}{*}{0.453} & -0.250 & \multirow{2}{*}{0.270} & 8.843 & \multirow{2}{*}{0.675} \\
$k_2$ & 3.349 & & 1.015 & & 0.471 & & 3.607 & & 0.948 & & 0.840 & \\
\midrule
$W_1$ & 0.711 & \multirow{6}{*}{0.750} & 0.360 & \multirow{6}{*}{0.384} & 0.201 & \multirow{6}{*}{0.322} & 0.691 & \multirow{6}{*}{0.755} & 0.406 & \multirow{6}{*}{0.424} & 0.242 & \multirow{6}{*}{0.350} \\
$W_2$ & 0.215 & & 0.140 & & 0.206 & & 0.278 & & 0.173 & & 0.197 & \\
$W_3$ & 0.144 & & 0.204 & & 0.169 & & 0.175 & & 0.239 & & 0.217 & \\
$W_4$ & 0.093 & & 0.058 & & 0.093 & & 0.121 & & 0.073 & & 0.092 & \\
$W_5$ & 0.643 & & 0.241 & & 0.115 & & 0.599 & & 0.269 & & 0.124 & \\
$W_6$ & 0.179 & & 0.111 & & 0.160 & & 0.225 & & 0.138 & & 0.150 & \\
\midrule
$f_1$ & -0.780 & \multirow{2}{*}{0.869} & -0.227 & \multirow{2}{*}{0.299} & -0.076 & \multirow{2}{*}{0.116} & -0.741 & \multirow{2}{*}{0.839} & -0.227 & \multirow{2}{*}{0.269} & -0.085 & \multirow{2}{*}{0.115} \\
$f_2$ & 0.073 & & 0.047 & & 0.039 & & 0.086 & & 0.052 & & 0.053 & \\
\midrule
$\text{RMSE}_{\text{train}}$ & 1.222 &  & 1.147 &  & 1.106 &  & 1.217 &  & 1.144 &  & 1.100 &  \\
$\text{RMSE}_{\text{test}}$ & 1.285 &  & 1.210 & & 1.165 &  & 1.283 &  & 1.205 &  & 1.158 &  \\
Runtime (hours) & 0.03 &  & 0.05 &  & 0.14 &  & 0.03 &  & 0.06 &  & 0.24 & \\
\bottomrule
\end{tabular}%
}
\caption{\label{tab:consolidated_results_R15}$LR = 15\%$. Bias and RMSE of the estimated parameters are reported for different numbers of time steps ($T$) and spatial locations ($m$). $RMSE_{\text{train}}$ and $RMSE_{\text{test}}$ denote the average prediction RMSEs computed on the training set $\mathcal{S}_i$ and the test set $\mathcal{S}_i^*$, respectively, across $M = 100$ Monte Carlo replications. Time indicates the average computation time over replications.}
\end{table}

Overall, as expected for the ML estimation, these results demonstrate that the proposed model yields consistent and increasingly accurate parameter estimates as more spatial and temporal information becomes available, confirming the robustness of the estimation procedure under increasing low-rank approximation. This means that, in general, the low-rank approximation can lead to a speed-up in the computation without losing a precise estimate. Furthermore, the estimate shows that the model performs well in the heterotopic case, meaning that it is able to estimate the cross-correlation among the observed processes. By evaluating the test RMSE under increasing levels of low-rank approximation $R$, while keeping $m$ fixed, we observe a slight increase in RMSE that is not statistically significant compared to the case $LR = 100\%$. This demonstrates that the proposed approximation procedure substantially reduces computation time while maintaining consistent parameter estimates and accurate spatial mapping performance. The computation time refers to a machine equipped with an Intel Xeon Platinum 8460Y+ CPU and 1007 GB RAM using Python 3.10.15.

\section{Analysis of a multivariate spatio-temporal air quality dataset} \label{sec:case_study}

We now illustrate how the model is used to analyse a large spatio-temporal dataset with multiple pollutants, heterogeneous spatial coverage, and substantial missingness.

We focus on two pollutants: coarse particulate matter with aerodynamic diameter less than 10~$\mu$m ($PM_{10}$) and fine particulate matter with diameter less than 2.5~$\mu$m ($PM_{2.5}$) in Italy. These pollutants are typically observed at partially different monitoring stations (heterotopic) and are known to be highly correlated.
\cite{fasso2016european} shows that the estimated correlation between co-located $PM_{10}$ and $PM_{2.5}$ at the European level is close to 0.93 after controlling for meteorology and orography. Hence, leveraging this correlation, the aim is to obtain maps of daily concentrations at the national level. This requires adopting a statistical model capable of handling the complexity of the bivariate pollutant phenomenon, while controlling for meteorology and orography, as well as accounting for an unbalanced monitoring network, heterotopic observations, and extensive missing data.

\subsection{Data description}
We use the AQCLIM-GRINS dataset provided by \cite{AQCLIM_GRINS}, which contains daily air quality and meteorological measurements across Italy, covering the five-and-a-half-year period from July 1, 2018, to December 30, 2023.  The AQCLIM-GRINS dataset extends the AGRIMONIA dataset \citep{fasso2023agrimonia} to the entire Italian territory and has been harmonised by the GRINS project (\url{https://www.grins.it/}, accessed 26/07/2025).

The particulate matter data consists of daily average concentrations (in $\mu g / m^3$) collected at 661 ground-level monitoring stations irregularly distributed across Italy. 
Also, we include the following covariates: Altitude (in metres above sea level), Temperature at 2 metres (in $^\circ$C), Relative Humidity (RH; in \%), and Wind Speed (in $m / s $), along with an intercept term. Table \ref{tab:aq_stats} summarises the variables used in this study and the associated main statistics.

\begin{figure}
    \centering
    \includegraphics[scale=0.5]{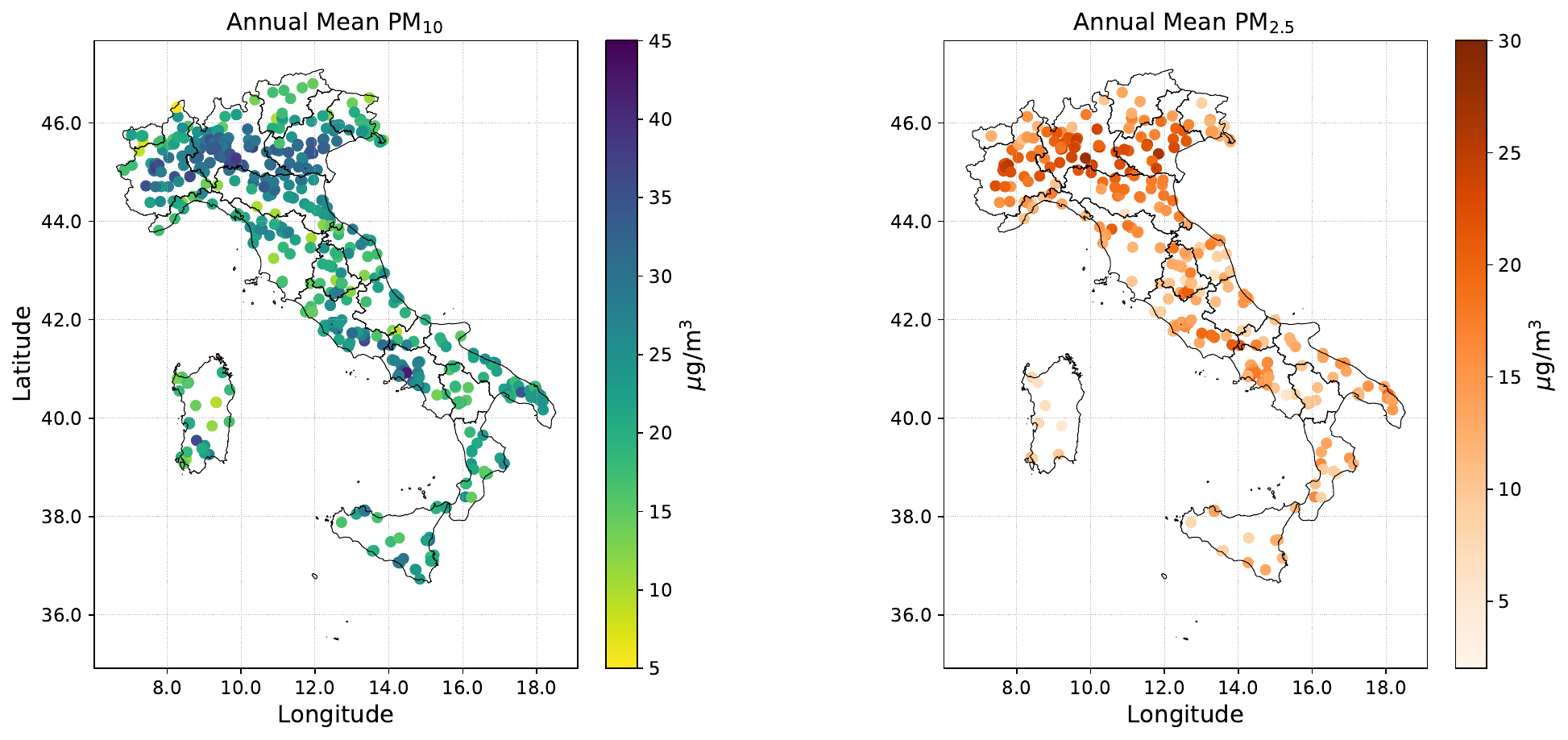}
    \caption{Monitoring network and overall averages of $PM_{10}$ (left) and $PM_{2.5}$(right) concentration over Italy. A total of $S = 661$ stations are available: $S_1 = 574$ stations measuring $PM_{10}$, and $S_2 = 313$ measuring $PM_{2.5}$, with 304 stations providing co-located measurements.}
    \label{fig:observed_points}
\end{figure}

It can be noticed that the resulting measurement network in Figure~\ref{fig:observed_points} is partially heterotopic, meaning the variables are observed at intersecting but non-identical sets of locations. The considered period spans $T=2009$ days, and the dataset includes approximately 1.15 million observations of $PM_{10}$ from 574 monitoring stations, with 28.7\% missing values, and about 0.6 million observations of $PM_{2.5}$ from 313 stations, with 34.1\% missing values. In the sequel, concentrations above the $99.9$ percentile are considered outliers and replaced by a missing value. 

\subsection{Preliminary analysis}
\begin{figure}
    \centering
    \includegraphics[scale=0.38]{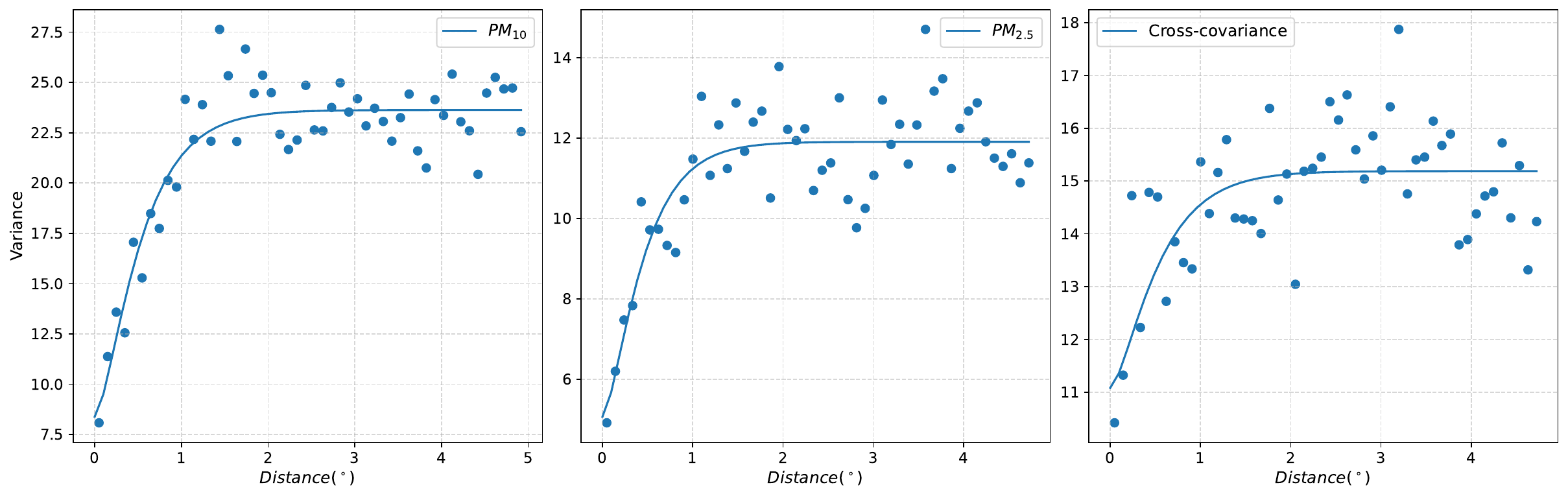}
    \caption{Spatial variograms and covariogram computed from the average concentration at each spatial location in $(\mu g/m^3)^2$. The plots show the variogram of $PM_{10}$ (left), the variogram of $PM_{2.5}$ (center), and the cross-covariogram between $PM_{10}$ and $PM_{2.5}$ (right). The fitted Matérn models are estimated with weighted least squares \citep{cressie1985fitting}. }
    \label{fig:emprical_variogram}
\end{figure}

Figure~\ref{fig:observed_points} highlights spatial patterns in pollutant average concentrations: nearby stations tend to record similar values, suggesting the presence of spatial autocorrelation and cross-correlation between the two particulate matter variables. 
This is confirmed by Figure \ref{fig:emprical_variogram} reporting the weighted least squares Matérn variograms and covariogram of station averages for $PM_{10}$ and $PM_{2.5}$ \citep{wackernagel2003multivariate,cressie1985fitting}. The variograms reveal a nugget effect, with estimated measurement error std $\hat{\sigma}_{i}$  
of 2.88 $\mu g / m^3$ for $PM_{10}$ and 2.24 $\mu g / m^3$ for $PM_{2.5}$. Accordingly, the model of the next section includes a measurement error component as specified in Equation~\eqref{eq:proposed_ssm2}. The estimated marginal variances of the spatial process are approximately 15.4 and 6.8 $(\mu g / m^3)^2$. As expected, the autocovariances of the two detrended processes decrease with distance. The estimated spatial ranges are about 0.37$^\circ$ (41 km) and 0.30$^\circ$ (33 km), respectively for $PM_{10}$ and $PM_{2.5}$. Similarly, the correlation between $PM_{10}$ and $PM_{2.5}$ is particularly high, see Figure \ref{fig:emprical_variogram} (right panel), and has a spatial range of 0.37$^\circ$ (41 km), which is close to the $PM_{10}$ behaviour since networks are almost overlapping.   

\begin{figure}
    \centering
    \includegraphics[scale=0.38]{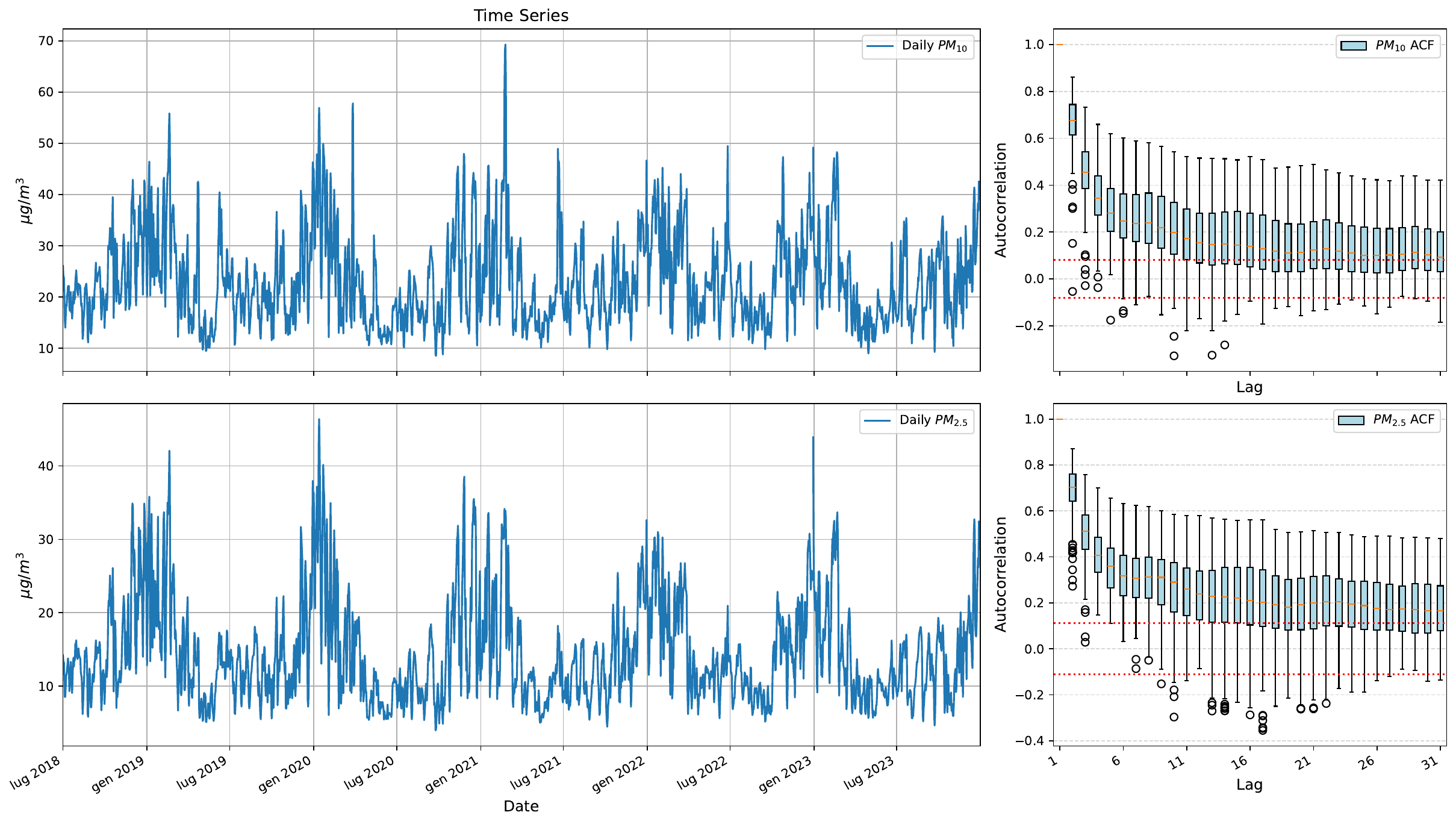}
    \caption{Temporal structure of particulate matter concentrations. The left column shows daily averages of $PM_{10}$ (top) and $PM_{2.5}$ (bottom); the right column displays corresponding autocorrelation functions (ACFs) over the first 30 lags, summarised across stations using boxplots.} 
\label{fig:timeseres}
\end{figure}

The temporal variation of the observed $PM_{10}$ and $PM_{2.5}$ concentrations is shown as daily averages among the sites in Figure \ref{fig:timeseres} (left panel). Not surprisingly, there is a clear seasonality with higher concentrations in winter. Moreover, the empirical autocorrelation function (ACF) shown in Figure \ref{fig:timeseres}. The first lag is close to 0.7 for both pollutants, indicating the presence of a temporal dependence.
Time series and ACF grouped by regions are summarised in the Supplementary material. 

\begin{table}[ht]
\caption{Summary statistics of variables from the AQCLIM-GRINS dataset across all locations.}
\label{tab:aq_stats}
\centering
\small
\renewcommand{\arraystretch}{0.55}
\begin{tabular}{lrrrrrr}
\toprule
\textbf{Statistics} & $PM_{10}$  & $PM_{2.5}$ & Altitude & Temperature  & RH\%  & Wind Speed \\
Unit of measurement
 & $\mu g/m^3$& $\mu g/m^3$& $m$ & $^\circ$C& & m/s \\
\midrule
\multicolumn{7}{l}{\emph{Percentiles}}\\
$\quad$ Minimum         & 0.00   & 0.00   & -2.00   & -24.68 & 15.27  & 0.11   \\
$\quad$ 5th         & 7.00   & 4.00   & 4.00    & 1.63   & 52.95  & 0.71   \\
$\quad$ 25th        & 13.76  & 8.00   & 26.00   & 8.31   & 66.59  & 1.18   \\
$\quad$ 50th        & 20.00  & 12.00  & 114.00  & 13.84  & 75.70  & 1.64   \\
$\quad$ 75th        & 29.55  & 19.00  & 271.00  & 20.29  & 83.63  & 2.45   \\
$\quad$ 95th        & 56.80  & 41.40  & 736.00  & 26.32  & 92.01  & 4.64   \\
$\quad$ Maximum         & 137.40 & 102.60 & 1770.00 & 35.75  & 99.98  & 16.54  \\
\multicolumn{7}{l}{\emph{Descriptive statistics}}\\
$\quad$ Mean        & 24.14  & 15.66  & 209.31  & 14.03  & 74.51  & 2.02   \\
$\quad$ Standard deviation     & 16.20  & 12.49  & 274.34  & 7.84   & 11.99  & 1.29   \\
$\quad$ Skewness    & 1.98   & 2.28   & 2.43    & -0.14  & -0.49  & 2.02   \\
$\quad$ Kurtosis    & 5.69   & 6.87   & 7.61    & -0.47  & -0.12  & 5.82   \\
\midrule
Number of stations $m$    & 574    & 313    & 661     & 661    & 661    & 661    \\
Missing proportion         & 0.29   & 0.34   & 0.00    & 0.00   & 0.00   & 0.00   \\
\bottomrule
\end{tabular}
\end{table}

\subsection{Application of the LR-SSM}

\begin{figure}
    \centering
    \includegraphics[scale=0.40]{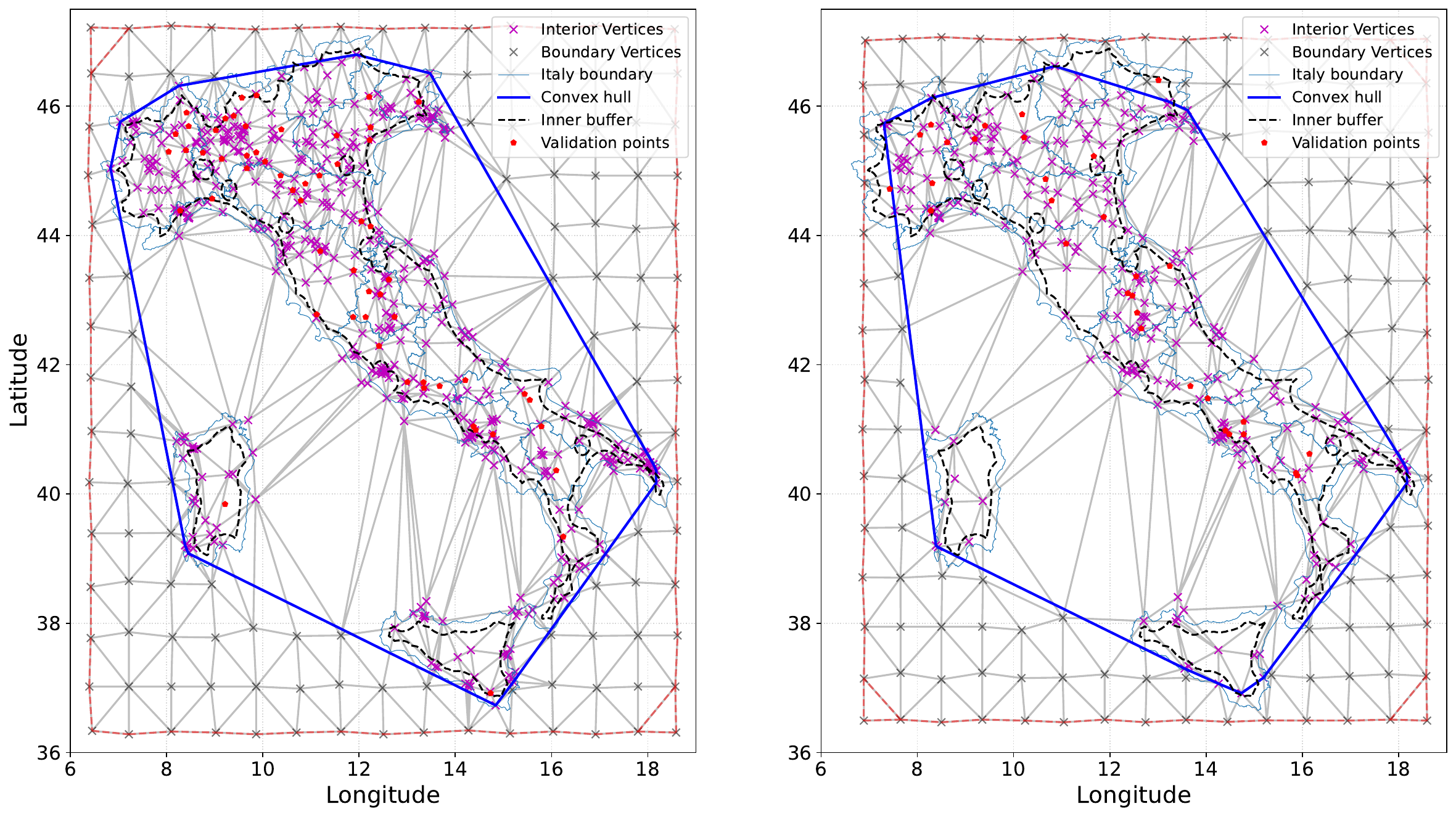}
    \caption{\label{fig:smoothed_domain} Spatial meshes generated via Laplacian smoothing. The left panel corresponds to the mesh constructed for the $PM_{10}$ dataset, and the right panel to the $PM_{2.5}$ dataset. Interior vertices (magenta), boundary vertices (green), convex hulls (blue), inner buffer (black), and validation locations (red) are shown. }
    
\end{figure}

To apply the proposed LR-SSM~\eqref{eq:proposed_ssm1}--\eqref{eq:proposed_ssm2} defined in Section~\ref{sec:model}, we discretise the considered domain using a Delaunay triangulation to form the spatial meshes $\mathcal{V}_R^j$ for $j = 1, 2$. This triangulation maximises the smallest angle among all triangles, promoting mesh regularity. To further ensure good mesh quality, we apply the Laplacian smoothing with a minimum internal angle threshold of 0.15$^\circ$. Additional details on the triangulation and smoothing procedures are given in Section~\ref{sec:triangulation}. The resulting discretised domain is shown in Figure~\ref{fig:smoothed_domain}.

We validate the model using a data split validation strategy tailored to the spatio-temporal setting \citep[see][for a review]{otto2024review}. Specifically, we adopt a leave-many-stations-out validation scheme. Basically, the entire time series of a subset of stations (the validation set) is withheld during training and later used to assess the model’s predictive performance. We construct the validation set by randomly selecting 10\% of the total stations of each pollutant, while the remaining 90\% are used for training. To avoid edge effects, we exclude stations close to the domain edge by selecting only stations located at least $0.15^\circ$ away from the boundary (see the inner buffer in Figure~\ref{fig:smoothed_domain}). The resulting training set includes 516 and 281 spatial points, while the validation set includes 58 and 32 spatial points for $PM_{10}$ and $PM_{2.5}$, respectively.  


To validate the low-rank approach and given a set of candidate R values, the mean square error optimality is of interest. Hence, considering $\hat{y}_i(\bm s, t)$ as in \eqref{eq:prediction_mean_lssm}, errors $e_i(\bm s, t) =  y_i(\bm s, t) - \hat{y}_i(\bm s, t)$ with $\bm s \in \Omega $ are computed and the LR-SSM candidates are compared in terms of RMSE and R$^2$ statistic, where, for each pollutant $i$, R$^2_i = 1 - \sum_{t = 1}^T \sum_{\bm s\in \mathcal{S}_{i,t}}(y_i(\bm s, t)  - \hat y_i(\bm s, t) )^2/\sum_{t = 1}^T \sum_{\bm s\in \mathcal{S}_{i,t}}(y_i(\bm s, t)  - \bar y_i )^2$, where $\bar y_i = T^{-1}\sum_{t = 1}^T |\mathcal{S}_{i,t}|^{-1} \sum_{\bm s\in \mathcal{S}_{i,t}}y_i(\bm s, t)$.  

We consider low-rank approximations where $LR$ is set to 100\%, 75\%, 50\%, and 25\%. Table \ref{tab:validation_results} summarises the validation results across models. 
As expected, for both pollutants, the training RMSE increases monotonically as the degree of approximation increases, while the test RMSE shows only a slight increase, at most by 6\% in the worst case. These results confirm that the low-rank model preserves predictive accuracy while significantly reducing computational time. Remarkably, even at the roughest approximation level (25\% of the rank), the validation RMSE increases by only 6\% for $PM_{10}$ and 15\% for $PM_{2.5}$ compared to the full-rank model, while computational time is reduced from 3h 37m to just 16m. This demonstrates that substantial efficiency gains can be achieved without compromising accuracy in a practically meaningful way. Notably, across all low-rank configurations, the test RMSEs remain comparable to those reported in \cite{rodeschini2024scenario} and \cite{fasso2016european}.


\begin{table}
\caption{\label{tab:validation_results}Performance and computational statistics of the estimation procedure across varying levels of low-rank approximation $LR$. The table reports the number of vertices ($R$) used in the approximation for each observed pollutant $PM_{10}$ and $PM_{2.5}$, RMSE and R$^2$ scores for both training and validation sets, as well as estimation time, and number of EM iterations. The computation time refers to a machine equipped with an Intel Xeon Platinum 8460Y+ CPU and 1007 GB RAM using Python 3.10.15.}
\renewcommand{\arraystretch}{0.65}
\begin{tabular}{ccc
                >{\centering\arraybackslash}p{1.2cm}
                >{\centering\arraybackslash}p{1.2cm}
                >{\centering\arraybackslash}p{1.2cm}
                >{\centering\arraybackslash}p{1.2cm}
                c
                c}
\toprule
\multirow{2}{*}{\textbf{LR}} & \multirow{2}{*}{$R$} & \multirow{2}{*}{\textbf{Variable}} 
& \multicolumn{2}{c}{\textbf{Train}} 
& \multicolumn{2}{c}{\textbf{Validation}} 
& \multirow{2}{*}{\textbf{Runtime}} 
& \multirow{2}{*}{\textbf{EM Iterations}} \\
\cmidrule(lr){4-5} \cmidrule(lr){6-7}
 & & & RMSE & R$^2$ & RMSE & R$^2$ & & \\
\midrule
\multirow{2}{*}{100\%} & 516 & $PM_{10}$  & 3.52 & 0.95 & 6.88 & 0.81 & \multirow{2}{*}{3h 37m} & \multirow{2}{*}{46} \\
                       & 281 & $PM_{2.5}$ & 2.24 & 0.96 & 5.44 & 0.80 &                        &                      \\ \hline
\multirow{2}{*}{75\%}  & 387 & $PM_{10}$  & 4.26 & 0.92 & 6.87 & 0.81 & \multirow{2}{*}{1h 51m} & \multirow{2}{*}{38} \\
                       & 211 & $PM_{2.5}$ & 2.94 & 0.93 & 5.47 & 0.80 &                        &                      \\ \hline
\multirow{2}{*}{50\%}  & 258 & $PM_{10}$  & 5.00 & 0.89 & 6.44 & 0.83 & \multirow{2}{*}{1h 00m} & \multirow{2}{*}{35} \\
                       & 141 & $PM_{2.5}$ & 3.51 & 0.90 & 5.62 & 0.79 &                        &                      \\ \hline
\multirow{2}{*}{25\%}  & 129 & $PM_{10}$  & 6.07 & 0.84 & 7.31 & 0.78 & \multirow{2}{*}{16m}    & \multirow{2}{*}{33} \\
                       & 71  & $PM_{2.5}$ & 4.43 & 0.84 & 6.26 & 0.74 &                        &                      \\ \hline
\end{tabular}
\end{table}

Table~\ref{table:empirical} reports the ML estimates and the associated uncertainty of the parameter vector \( \hat{\Pi}_{MLE} \) for different low-rank approximation levels. For both $PM_{10}$ and $PM_{2.5}$, altitude exhibits a small negative association, indicating that higher elevation is associated with a lower pollution levels. Temperature shows a negative effect, especially for $PM_{2.5}$. Relative Humidity has a positive correlation with both pollutants, more pronounced for $PM_{10}$. Wind Speed has a substantial negative effect on both processes, with higher winds contributing to the dispersion of pollutants. The results are coherent with the analysis of \cite{rodeschini2024scenario}.


\begin{table}[]
\caption{\label{table:empirical}ML estimates of regression coefficients and random effect parameters for the bivariate model of $PM_{10}$ and $PM_{2.5}$, for different low-rank approximation levels (LR = 100\%, 75\%, 50\% and 25\%). Covariates include Altitude, Temperature at 2 metres, RH, and Wind Speed. Values are followed by their standard deviations in parentheses.. 
}
\centering
\renewcommand{\arraystretch}{0.55}
\resizebox{\textwidth}{!}{
\begin{tabular}{l  rr|rr|rr|rr}
\toprule
                                 & \multicolumn{2}{c}{$LR = 100\%$}                & \multicolumn{2}{c}{$LR = 75\%$}                 & \multicolumn{2}{c}{$LR = 50\%$}                 & \multicolumn{2}{c}{$LR = 25\%$}                 \\
Parameter                        & \multicolumn{1}{c}{$PM_{10}$}   & \multicolumn{1}{c}{$PM_{2.5}$} & \multicolumn{1}{c}{$PM_{10}$}   & \multicolumn{1}{c}{$PM_{2.5}$} & \multicolumn{1}{c}{$PM_{10}$}   & \multicolumn{1}{c}{$PM_{2.5}$} & \multicolumn{1}{c}{$PM_{10}$}   & \multicolumn{1}{c}{$PM_{2.5}$} \\
\midrule
Intercept    & 15.762 & 12.154 & 16.809 & 12.694 & 16.814 & 13.145 & 15.220 & 12.673 \\
             & (0.029) & (0.026) & (0.034) & (0.032) & (0.038) & (0.036) & (0.044) & (0.044) \\
Altitude     & -0.012 & -0.007 & -0.014 & -0.008 & -0.012 & -0.007 & -0.012 & -0.009 \\
             & ($<$0.001) & ($<$0.001) & ($<$0.001) & ($<$0.001) & ($<$0.001) & ($<$0.001) & ($<$0.001) & ($<$0.001) \\
Temperature  & -0.055 & -0.236 & -0.060 & -0.239 & -0.056 & -0.257 & -0.053 & -0.269 \\
             & ($<$0.001) & ($<$0.001) & (0.001) & (0.001) & (0.001) & (0.001) & (0.001) & (0.001) \\
RH           & 0.088 & 0.062 & 0.082 & 0.057 & 0.074 & 0.054 & 0.096 & 0.069 \\
             & ($<$0.001) & ($<$0.001) & ($<$0.001) & ($<$0.001) & ($<$0.001) & ($<$0.001) & ($<$0.001) & ($<$0.001) \\
Wind Speed   & -0.760 & -0.874 & -0.854 & -0.863 & -0.788 & -0.830 & -0.837 & -0.926 \\
             & (0.003) & (0.003) & (0.003) & (0.003) & (0.004) & (0.004) & (0.004) & (0.004) \\
W$_1$        & 6.998 & 2.439 & 6.515 & 2.575 & 6.200 & 2.752 & 6.008 & 3.375 \\
             & (0.001) & (0.001) & (0.001) & (0.001) & (0.002) & (0.001) & (0.002) & (0.001) \\
W$_2$        & 3.467 & 3.688 & 3.351 & 3.356 & 3.425 & 3.198 & 3.584 & 3.184 \\
             & (0.002) & (0.001) & (0.002) & (0.001) & (0.002) & (0.001) & (0.002) & (0.002) \\
$\sigma^2_1$ & 18.535 & 7.825 & 24.642 & 11.633 & 30.808 & 15.107 & 41.300 & 21.828 \\
             & (0.008) & (0.004) & (0.012) & (0.008) & (0.016) & (0.011) & (0.025) & (0.020) \\
$f$          & 0.823 & 0.869 & 0.827 & 0.889 & 0.844 & 0.903 & 0.895 & 0.946 \\
             & ($<$0.001) & ($<$0.001) & ($<$0.001) & ($<$0.001) & ($<$0.001) & ($<$0.001) & ($<$0.001) & ($<$0.001) \\
$k$          & 3.349 & 3.232 & 2.676 & 2.197 & 2.118 & 1.844 & 1.578 & 1.904 \\
             & (0.002) & (0.003) & (0.002) & (0.002) & (0.002) & (0.003) & (0.002) & (0.004) \\

\bottomrule
\end{tabular}
}
\end{table}

\begin{figure}
    \centering
    \includegraphics[scale=0.70]{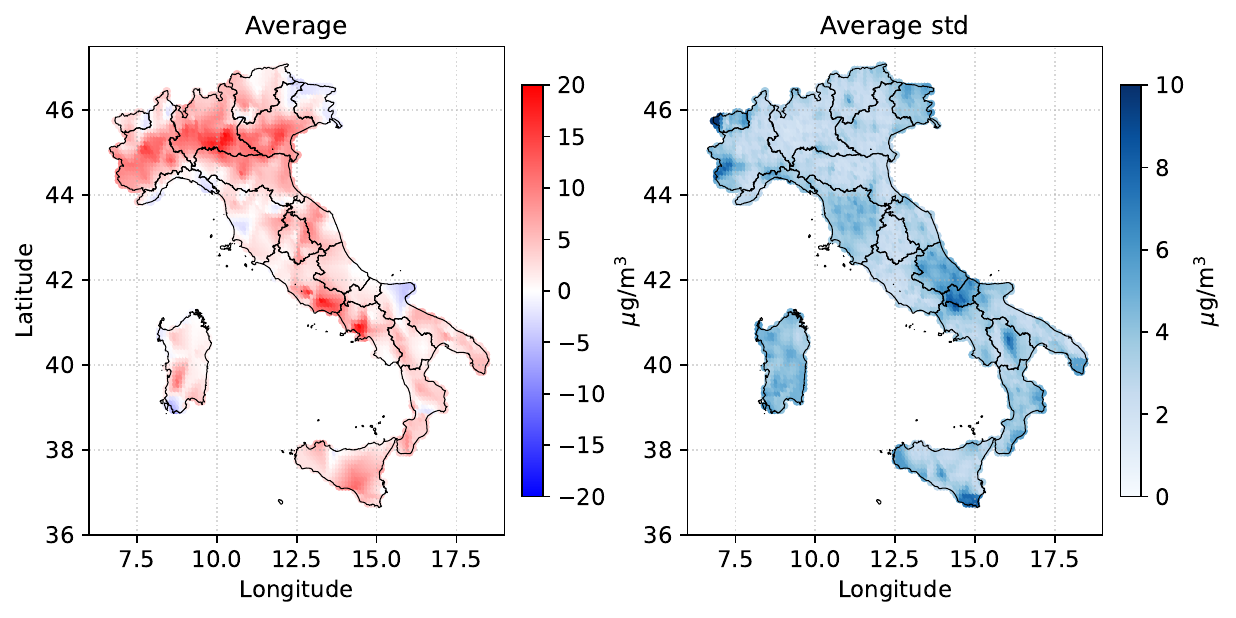}
    \caption{Estimated average latent effect component  on $PM_{10}$ (left) and associated with the average standard deviation (right).}
    \label{fig:average_maps_pm10_100}
\end{figure}

\begin{figure}
    \centering
    \includegraphics[scale=0.55]{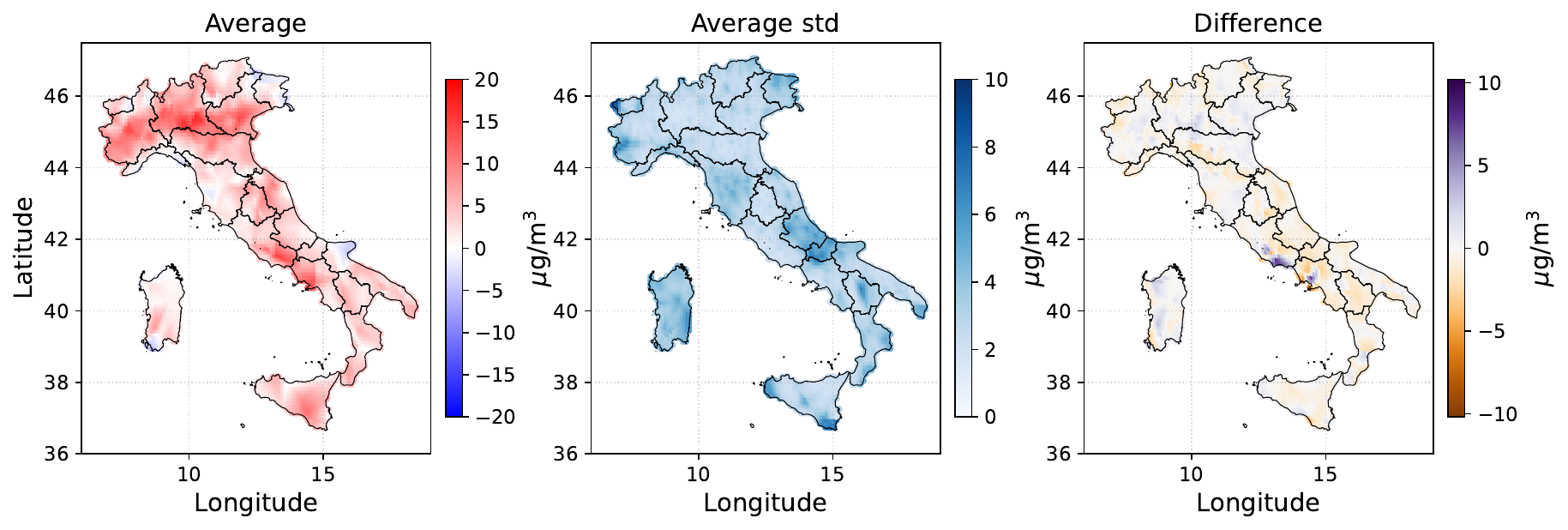}
    \caption{$LR=75\%$. Estimated average latent effect component  on $PM_{10}$ (left),  associated with the average standard deviation (right) and average difference between $LR=100\%$ and $LR=75\%$ estimations (right). }
    \label{fig:average_maps_pm10_75}
\end{figure}

\begin{figure}
    \centering
    \includegraphics[scale=0.55]{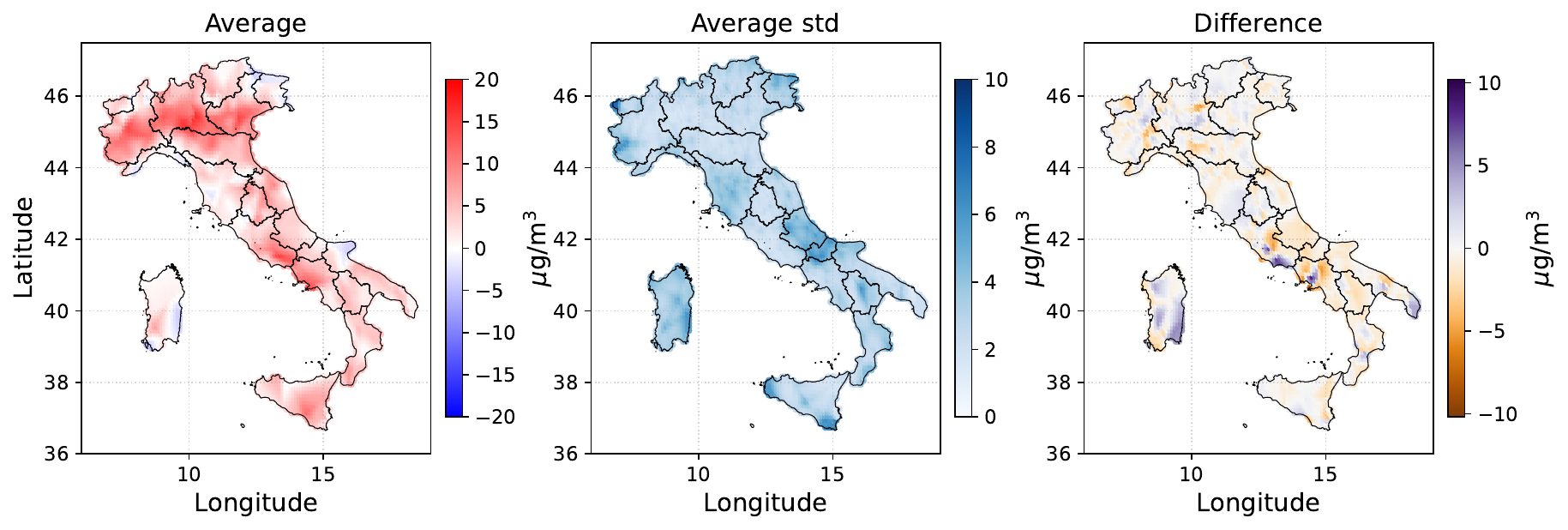}
    \caption{$LR=50\%$. Estimated average latent effect component  on $PM_{10}$ (left),  associated with the average standard deviation (right) and average difference between $LR=100\%$ and $LR=50\%$ estimations (right).}
    \label{fig:average_maps_pm10_50}
\end{figure}

\begin{figure}
    \centering
    \includegraphics[scale=0.55]{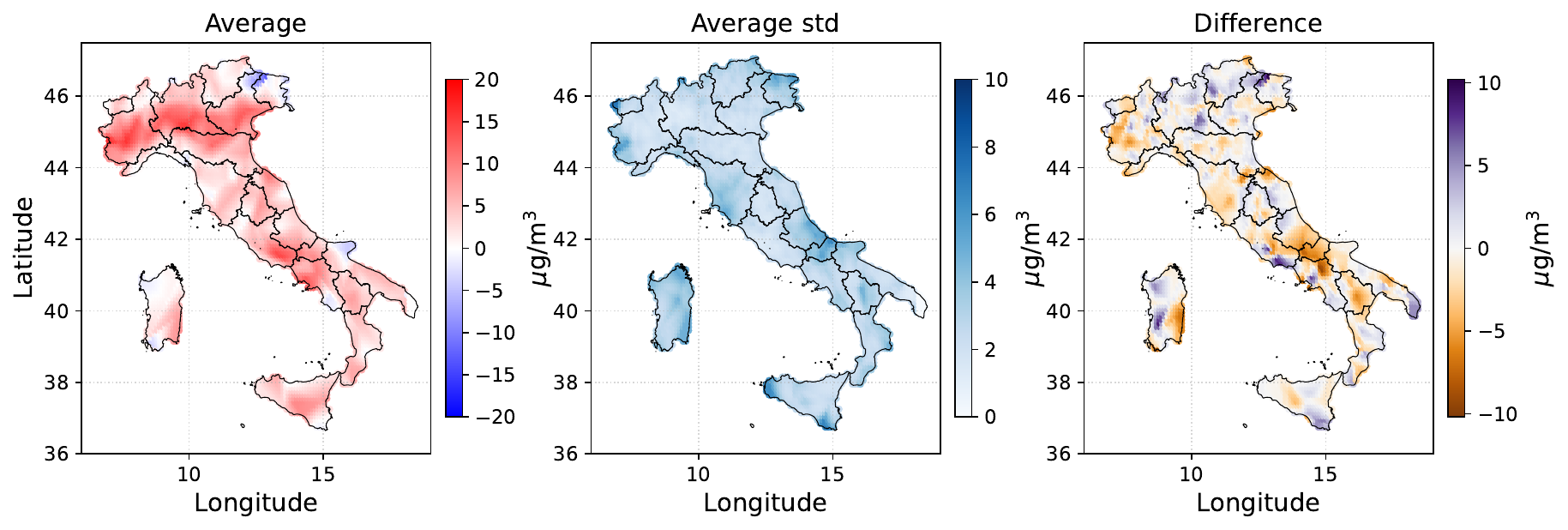}
    \caption{$LR=25\%$. Estimated average latent effect component  on $PM_{10}$ (left),  associated with the average standard deviation (right) and average difference between $LR=100\%$ and $LR=25\%$ estimations (right).}
    \label{fig:average_maps_pm10_25}
\end{figure}

Finally, Figures \ref{fig:average_maps_pm10_100}, \ref{fig:average_maps_pm10_75}, \ref{fig:average_maps_pm10_50}, and \ref{fig:average_maps_pm10_25} show the estimated average random effect component of the $PM_{10}$ across the study domain. Each map depicts the mean over all available years, corresponding to the average over time of $\hat{\bm W}\bm H(\bm s)\hat{\bm z}_t$ and its standard deviation, as defined in the predictive equations \eqref{eq:prediction_mean_lssm} and \eqref{eq:prediction_std_lssm}. As expected, prediction performance decreases as the approximation level increases. Nevertheless, the average standard deviation remains low across all cases. The magnitude of the random effects suggests some level of model misspecification, specifically, that in certain areas, the available covariates are insufficient to fully capture the variability of the process. In Figure \ref{fig:average_maps_pm10_100}, the Po Valley in Northern Italy, one of the most polluted regions in Europe, is clearly distinguishable due to its large positive random effects. This result aligns with findings from other studies focusing on Northern Italy, such as \cite{rodeschini2024scenario}. 
As the approximation level increases ($LR$ decreases), the latent process becomes smoother, which is also reflected in the reduced values of the estimated rescaling parameters $k_i$. In the approximation case $LR = 25\%$ in Figure \ref{fig:average_maps_pm10_25}, some local variability present in the data is lost. Indeed, the difference from the full-rank case (i.e., $LR = 100\%$) is especially noticeable in regions with a high density of stations but where the mesh is relatively sparse. Regarding the estimated uncertainty, it remains consistent with the test RMSE, which is approximately 6 $\mu g / m^3$. Notably, no boundary effects are observed, and the uncertainty remains stable even near the edges of the domain.

\section{Conclusion}\label{sec:conclusion}

This paper develops a low-rank multivariate state–space model in which each spatial innovation is represented through the SPDE–GMRF construction on a finite–element mesh.  By embedding this sparse representation inside the Kalman filter and coupling it with an EM algorithm that supplies mostly closed-form updates, we obtained a likelihood-based estimator whose computational cost grows only linearly with the number of basis functions.  Theoretical analysis showed weak convergence of the approximated latent field to the target Matérn field as the mesh is refined, and it established a uniform, time-independent bound on the resulting prediction error.  Numerical experiments demonstrated that, in practice, reducing the rank can often achieve linear SSM accuracy. The air-quality application showed the method’s ability to handle large, irregular datasets with substantial missingness. 

Several directions naturally follow from this work. Because the methodology is based on the SPDE-GMRF approach, the entire framework can be transferred from planar regions to curved two-dimensional manifolds such as the sphere, or even to higher-dimensional Riemannian spaces, by adapting the mesh and basis functions to the underlying geometry.  Likewise, substituting the continuous domain with a discrete graph, it would be possible to extend the model to networks such as road systems, power grids or social graphs \citep{porcu2023stationary}; the resulting precision matrix remains sparse, so the linear-time Kalman–EM machinery is preserved.  A further refinement could allow the mesh to evolve in time, thereby allocating resolution where the signal is most variable without inflating the overall computational budget. A comparison with the current implementation of the SSM for spatio-temporal data will be pursued in future research.  

In summary, the proposed approach retains the physical interpretability of SPDE models, the temporal elegance of the state–space formulation, and the scalability of low-rank methods.  Its extension to manifolds and graphs appears not only straightforward but also highly advantageous, promising fast, principled inference with quantified uncertainty for an even broader class of spatio-temporal problems.

\section*{Supplementary Materials} 
A supplement to the main article, including additional simulated results, is provided in the supplementary materials. (.pdf file). 

The Python code (.py script) and instructions for reproducing the results in the article are available at the GitHub repository: \url{https://github.com/jacopoRodeschini/Low_Rank_State_Space_Model}

\section*{Acknowledgment}
This work is partially funded by the National Recovery and Resilience Plan (NRRP), Mission 4 Component 2 Investment 1.3, under Call for tender No. 341 of 15/03/2022, funded by the Italian Ministry of University and Research and the European Union’s NextGenerationEU initiative. Award Number: PE 00000018, Concession Decree No. 1558 of 11/10/2022, CUP F83C22001720001, under the project titled Growing Resilient Inclusive and Sustainable (GRINS). 
Lorenzo Tedesco is also a member of the GNCS group of INdAM, whose support is gratefully acknowledged. This research was partially supported by the German Research Foundation, project number 501539976 (Philipp Otto).

\newpage
\appendix
\section*{Appendix}
This appendix collects supporting material for the main text. Section~\ref{appendix:notation} defines the notation used throughout the appendix. Section~\ref{appendix:identification} provides the detailed proof of the identifiability theorem. Section~\ref{appendix:weak_approxiation} discusses the weak convergence properties of the low-rank approximation to the latent process, with full theoretical justifications. Section~\ref{appendix:EM} presents the derivation and implementation details of the EM algorithm used for inference. Finally, Section~\ref{appendix:boundary} explores the role of Neumann boundary conditions in the SPDE approximation and their impact on the induced covariance structure.

\section{Notation}\label{appendix:notation}
We introduce the following notation. For two matrices \( A \) and \( B \) of dimensions \( r_A \times c \) and \( r_B \times c \), respectively, we denote by \( [A; B] \) the \( (r_A + r_B) \times c \) matrix obtained by stacking \( A \) on top of \( B \). We use $A \otimes B $ to denote the Kronecker product. Analogously, for two matrices \( A \) and \( B \) with dimensions \( r \times c_A \) and \( r \times c_B \), respectively, we define the column-wise juxtaposition of \( A \) and \( B \), denoted by \( [A, B] \), as the \( r \times (c_A + c_B) \) matrix obtained by placing \( A \) to the left of \( B \). Moreover, for matrices \( A \) and \( B \) of dimensions \( r_A \times c_A \) and \( r_B \times c_B \), respectively, we denote by \( A \oplus B \) the block diagonal matrix of dimension \( (r_A + r_B) \times (c_A + c_B) \), constructed with \( A \) and \( B \) placed as diagonal blocks and zeros elsewhere. Lastly, denote by $|A|$ the determinant of a square matrix $A$ and by $\bm I_n$ the identity matrix of dimension $n$.

\section{Proof of Theorem~\ref{thm:identifiability}}\label{appendix:identification}

\begin{proof}
\textit{Proof} Assume two parameter sets
\[
\Pi=(\bm\beta,\bm\sigma^{2},\bm f,\bm W,\{k_i\}_{i=1}^{q}),
\qquad
\Pi^{\star}=(\bm\beta^{\star},\bm\sigma^{2\star},\bm f^{\star},\bm W^{\star},\{k_i^{\star}\}_{i=1}^{q})
\]
generate the same joint distribution for the observable field
$\{\bm y(\bm s,t):(\bm s,t)\in\mathcal S\times\mathbb N\}$ under the
LR-SSM~\eqref{eq:proposed_ssm1}–\eqref{eq:proposed_ssm2} for a fixed valueof $R$. 
As the latent state sequence 
$\{\bm z(\mathcal V_R,t)\}_{t\in\mathbb N}$ is strictly stationary,
ergodic and has mean zero, the Strong Law of Large Numbers yields
\[
\frac1T\sum_{t=1}^{T}\bm z(\mathcal V_R,t)
      \;\xrightarrow{\text{a.s.}}\; \bm 0\quad (T\to\infty).
\]
Define $L_{\bm X} = \lim_{T\to \infty} \frac{1}{T} \int \bm X(\bm s,t)d\bm s$ which exists finite by hypothesis. Then, 
$$
\bm \beta = (L_{\bm X}L_{\bm X}')^{-1}L_{\bm X}'\lim_{T\to \infty} \frac{1}{T} \int y(\bm s,t)d\bm s = \bm \beta^\star.
$$

Centre the observations with
\(
\bm r(\bm s,t)=\bm y(\bm s,t)-\bm X'(\bm s,t)\bm\beta
              =\bm W\bm\Psi_R(\bm s)\bm z_t+\bm\varepsilon(\bm s,t),
\)
where
$\bm z_t=\bm z(\mathcal V_R,t)$ obeys
\(
\bm z_t=(\operatorname{diag}(\bm f)\otimes\bm I_R)\bm z_{t-1}+\bm\eta_t.
\)

Let
\(
\bm\Gamma_1(\bm s,\bm s')
   =\operatorname{Cov}\bigl(\bm r(\bm s,t),\bm r(\bm s',t-1)\bigr).
\)
Using stationarity of $\bm z_t$ with covariance
\(
\bm V_{\bm z}
   =\oplus_{i=1}^q(1-f_i^2)^{-1}\bm Q_{\kappa_i}^{-1},
\)
\[
\bm\Gamma_1(\bm s,\bm s')
   =\bm W\bm\Psi_R(\bm s)
     \bigl[\operatorname{diag}(\bm f)\otimes\bm I_R\bigr]
     \bm V_{\bm z}
     \bm\Psi_R(\bm s')'\bm W'.
\]
Define
\[
\bm H=
\Bigl(\!\!\int_{\mathcal S^2}\!\bm\Psi_R'(\bm s)\bm W^{\dagger}
        \bm\Gamma_0(\bm s,\bm s')\bm W^{\dagger\prime}\bm\Psi_R(\bm s')
        \,d\bm s\,d\bm s'\Bigr)^{-1}
\!\!
\int_{\mathcal S^2}\!\bm\Psi_R'(\bm s)\bm W^{\dagger}
        \bm\Gamma_1(\bm s,\bm s')\bm W^{\dagger\prime}\bm\Psi_R(\bm s')
        \,d\bm s\,d\bm s',
\]
with $\bm W^{\dagger}$ any right inverse of $\bm W$.
The integrals average over a dense spatial grid, so
$\bm H\to\operatorname{diag}(\bm f)$.
Equality of distributions implies the same limit for~$\Pi^{\star}$, and
the strict ordering $f_1>\dots>f_q$
(Assumption~\ref{ass:identifiability}\,(ii)) forces
$\bm f=\bm f^{\star}$.

Stacking a dense set of locations gives
\[
\operatorname{Cov}(\bm r_t)
   =(\bm I_m\otimes\bm W)
     (\bm P\bm V_{\bm z}\bm P')
     (\bm I_m\otimes\bm W)'
     +\bm I_m\otimes\bm\Sigma,
\]
where $\bm P$ stacks the basis evaluations.
Because $\bm P$ and $\bm W$ have full column rank
(Assumption~\ref{ass:identifiability}\,(iii)–(iv)),
standard factor-analysis arguments \citep{bai2015identification},
together with the column-scaling convention, yield
$\bm W=\bm W^{\star}$ and $\bm V_{\bm z}=\bm V_{\bm z}^{\star}$.
The SPDE construction gives a one-to-one map
$k_i\mapsto\bm Q_{k_i}$, so $k_i=k_i^{\star}$ for all~$i$.

From the diagonal of
$\operatorname{Var}\bigl(\bm r(\bm s,t)\bigr)$,
every element of the diagonal matrix $\bm\Sigma$ is determined; hence
$\bm\sigma^{2}=\bm\sigma^{2\star}$. All components coincide, so $\Pi=\Pi^{\star}$.\hfill$\square$
\end{proof}

\section{Weak Convergence of the Low-Rank Approximation}\label{appendix:weak_approxiation}

In this section, we discuss the weak convergence of the process $\bm \Psi_R(\bm s)\bm z(\mathcal{V}_R, t)$ to the continuous process $\bm z(\bm s, t)$. To this end, we first introduce the concept of Gaussian Markov Random Fields (GMRFs). 

Let $G = (\mathcal{V}, \mathcal{E})$ be a finite undirected graph, where $\mathcal{V}$ is a set of vertices with cardinality $R < \infty$, and $\mathcal{E}$ is the set of edges. Let $\mathcal{X} = \{\mathcal{X}_v\}_{v \in \mathcal{V}}$ denote a collection of random variables indexed by $\mathcal{V}$. We say that $\mathcal{X}$ forms a zero-mean Markov Random Field defined by $G$ if it satisfies the local Markov property:
\begin{equation}\label{eq:MGRFs}
    \mathcal{X}_i \independent \mathcal{X}_{\mathcal{V} \setminus \partial i} \mid \mathcal{X}_{\partial i}, \quad \forall i \in \mathcal{V},
\end{equation}
where $\partial i$ denotes the set of neighbours of node $i$, i.e., all vertices adjacent to $i$ in the graph. Specifically, $\mathcal{X}$ is a GMRF on $G$ if it is Gaussian with zero mean and precision matrix $Q$ such that $Q_{ij} = 0$ whenever $j \notin \{\partial i \cup i\}$. This sparsity in the precision matrix facilitates the use of efficient numerical methods.

The explicit link between Gaussian Field (GF) parameters and GMRF parameters is established in \cite{lindgren2011explicit}. The goal is to approximate a GF with covariance matrix $\Sigma$ using a GMRF with precision matrix $Q$ such that $Q^{-1}$ approximates $\Sigma$ in a suitable norm. A GMRF representation of a Matérn GF can be efficiently constructed using the stochastic partial differential equation (SPDE) approach, in combination with the Finite Element Method (FEM) over a triangulated domain.

Let $\Omega \subset \mathbb{R}^d$ be a bounded domain with a smooth boundary $\partial \Omega \in C^2(\mathbb{R}^d)$. As shown in \cite{guttorp2006studies}, a one-dimensional GF \( x(\bm s) \) with Matérn covariance characterised by parameters $(\sigma^2, \nu, \kappa)$ is a stationary solution of:
\begin{equation}\label{eq:partial_differential_equation}
(\kappa^2 - \Delta)^{\alpha/2} x(\bm s) = \mathcal{W}(\bm s), \quad \bm s \in \Omega, \quad \kappa > 0,
\end{equation}
where $\Delta = \sum_{i=1}^{d} \frac{\partial^2}{\partial x_i^2}$ is the Laplacian and $\alpha = \nu + d/2$. The operator $(\kappa^2 - \Delta)^{\alpha/2}$ is defined via its spectral properties, as described in \cite{whittle1954stationary,whittle1963stochastic}. Here, $\mathcal{W}$ denotes spatial Gaussian white noise with unit variance, and $x(\bm s)$ has marginal variance:
\[
\sigma^2 = \frac{\Gamma(\nu)}{\Gamma(\nu + d/2)(4\pi)^{d/2} \kappa^{2\nu}}.
\]

For simplicity, we consider the common spatial case with $d = 2$ and $\alpha = 2$, implying $\sigma^2 = (4\pi \kappa^{2})^{-1}$. The spectral density of the stationary solution to \eqref{eq:partial_differential_equation} is given by:
\[
R(\bm k) = (2\pi)^{-2}(\kappa^2 + \|\bm k\|^2)^{-2}.
\]
Using the Fourier transform, the fractional Laplacian is defined as:
\begin{equation}\label{eq:fourier_spectral}
\mathcal{F} \{ (\kappa^2 - \Delta)\varphi \}(\bm k) = (\kappa^2 + \|\bm k\|^2) \mathcal{F}\varphi(\bm k),
\end{equation}
where $\varphi$ is any function on $\mathbb{R}^d$ with an invertible right-hand side under the Fourier transform. To avoid null-space solutions, Neumann boundary conditions (zero normal derivatives on $\partial \Omega$) are imposed.

To numerically solve \eqref{eq:partial_differential_equation}, we apply FEM \citep{ciarlet2002finite,brenner2008mathematical,quarteroni2008numerical}. Let $\Omega$ be partitioned into non-overlapping triangles via Delaunay triangulation, resulting in a mesh graph $G = (\mathcal{V}, \mathcal{E})$ with $R = |\mathcal{V}|$.

The stochastic weak formulation of the SPDE requires defining the inner product:
\[
\langle f, g \rangle = \int_{\Omega} f(\bm s) g(\bm s) \, d\bm s.
\]
The weak solution satisfies:
\begin{equation}\label{eq:SPDE_distribution_equality}
\langle \phi_j, (\kappa^2 - \Delta) x \rangle \overset{w}{=} \langle \phi_j, \mathcal{W} \rangle, \quad \forall j,
\end{equation}
for suitable test functions $\{\phi_j\}_{j=1}^R$. We represent the FEM approximation as:
\begin{equation}\label{eq:FEM_approximation}
x^R(\bm s) = \sum_{k=1}^{R} \psi_k(\bm s) w_k = \bm \psi_R' \bm w,
\end{equation}
where $\bm \psi_R(\bm s) = (\psi_1(\bm s), \dots, \psi_R(\bm s))'$ are piecewise linear basis functions, and $\bm w$ are Gaussian weights. Boundary conditions are enforced by choosing basis functions with zero normal derivatives on $\partial \Omega$, ensuring solutions reside in the Hilbert space $\mathcal{H}_R(\Omega)$.

Choosing $\phi_k = \psi_k$ yields the Galerkin approximation. Define matrices:
\begin{align*}
\bm C_{ij} &= \langle \psi_i, \psi_j \rangle,\\
\bm G_{ij} &= \langle \nabla \psi_i, \nabla \psi_j \rangle,\\
\bm K_{\kappa^2} &= \kappa^2 \bm C + \bm G.
\end{align*}
Under Neumann conditions, the precision matrix for $\bm w$ becomes:
\begin{align}\label{eq:Q}
    \bm Q_{\kappa^2} = \bm K_{\kappa^2} \bm C^{-1} \bm K_{\kappa^2}.
\end{align}
While $\bm C^{-1}$ is dense, one can replace $\bm C$ with a diagonal matrix $\tilde{\bm C}$, where $\tilde{\bm C}_{ii} = \langle \psi_i, 1 \rangle$, to obtain a sparse $\bm Q$ \citep{lindgren2011explicit}.

As shown in equation (11) of \cite{lindgren2011explicit}, the approximation error is bounded as follows:
\begin{equation}\label{eq:consistency}
\sup_{f \in \mathcal{H}^1, \|f\|_{\mathcal{H}^1} \leq 1} \mathbb{E}\left[ \langle f, x^R - x \rangle_{\mathcal{H}^1}^2 \right] \leq c h^2,
\end{equation}
where $\mathcal{H}^1(\Omega, \kappa)$ is the Hilbert space:
\[
\langle \phi, \psi \rangle_{\mathcal{H}^1(\Omega, \kappa)} = \kappa^2 \langle \phi, \psi \rangle + \langle \nabla \phi, \nabla \psi \rangle,
\]
and $h$ denotes the mesh resolution.

We consider the following convergence definition,as formulated in \cite{lindgren2011explicit}. Let $x^R$ be a sequence of $L^2(\Omega)$-bounded Gaussian fields. We say $x^R \xrightarrow{D\{L^2(\Omega)\}} x$ if for all $f, g \in L^2(\Omega)$,
\begin{align*}
\mathbb{E}[\langle f, x^R \rangle] \to \mathbb{E}[\langle f, x \rangle], \quad \text{and}\quad 
\operatorname{Cov}(\langle f, x^R \rangle, \langle g, x^R \rangle) \to \operatorname{Cov}(\langle f, x \rangle, \langle g, x \rangle).
\end{align*}

\begin{theorem}\label{theo:convergence_FEM}
Let $x$ be a weak solution to the SPDE $\mathcal{L}x = \mathcal{W}$ with $\mathcal{L} = (\kappa^2 - \Delta)$ and Neumann boundary conditions. Let $x^R$ be its FEM approximation in $\mathcal{H}^1_R(\Omega, \kappa)$ using Gaussian white noise $\mathcal{W}$. Then,
\begin{align*}
x^R &\xrightarrow{D\{L^2(\Omega)\}} x,\\
\mathcal{L}x^R &\xrightarrow{D\{L^2(\Omega)\}} \mathcal{L}x,
\end{align*}
for $R\to \infty$ provided that $\bigcup_R \mathcal{H}^1_R(\Omega, \kappa)$ is dense in $\mathcal{H}^1(\Omega, \kappa)$ and $x^R$ is the Galerkin solution.
\end{theorem}
The proof is available in \cite{lindgren2011explicit}.

\subsection{Proof of Theorem \ref{theo:z_convergeces}}\label{appendix:proof_z_convergence}
\begin{proof}
By Theorem~\ref{theo:convergence_FEM}, and fixing the component $i$ of the process $\bm \eta(\cdot, t)$ model, we have $\eta_i(\cdot, t)^R \xrightarrow{D\{L^2(\Omega)\}}\eta_i(\cdot,t)$ where $\eta_i^R(\cdot, t) = \bm \psi_R(s)'\bm \eta_i(\mathcal{V}_R, t)$ for some Gaussian weights $\bm \eta_i(\mathcal{V}_R, t)$, for each $t = 0,\dots, T$. Due to independence, and the fact that $t =1,\dots, T$ varies over a finite set, we have that $\eta_i^R(\cdot, \cdot)\xrightarrow{D\{L^2(\Omega^T)\}}\eta_i(\cdot, \cdot)$. Now, due to SSM \eqref{eq:proposed_ssm1}-\eqref{eq:proposed_ssm2}, we have that $z_i(\cdot,t) = \sum_{j=0}^tf_i^j\eta_i(\cdot, j)$. Thus, by the linearity of the approximation $\eta_i^R(\cdot, t)$, we have an immediate approximation of $z_i(\cdot, \cdot)$ as  $z_i^R(\bm s,t) = \bm \psi_R(\bm s) \bm z_i(\mathcal{V}_R, t)$ where the Gaussian weights are $ \bm z_i(\mathcal{V}_R, t) = \sum_{j=0}^Tf_i^{t-j}\bm \eta_i(\mathcal{V}_R, t)$. Therefore, $\eta_i^R(\cdot, \cdot)\xrightarrow{D\{L^2(\Omega^T)\}}\eta_i(\cdot, \cdot)$ implies $z_i^R(\cdot, \cdot)\xrightarrow{D\{L^2(\Omega^T)\}}z_i(\cdot, \cdot)$. As the number of components is finite $i=1,\dots, q$, we obtain the aimed result: 
\[
\bm z^R(\bm s, t) = \bm \Psi_R(\bm s)\bm z(\mathcal{V}_R, t) \xrightarrow{D\{L^2(\Omega^{qT})\}} \bm z(\bm s, t).
\]
This convergence result also implies that \begin{align*}
\mathbb{E}[\bm \Psi_R(\bm s)\bm z(\mathcal{V}_R, t)] &\to \mathbb{E}[\bm z(\bm s, t)], \\
\text{Cov}(\bm \Psi_R(\bm s)\bm z(\mathcal{V}_R, t), \bm \Psi_R(\tilde{\bm s})\bm z(\mathcal{V}_R, \tilde{t})) &\to \text{Cov}(\bm z(\bm s, t), \bm z(\tilde{\bm s}, \tilde{t})).
\end{align*}
$\,\hfill\square$
\end{proof}

\subsection{Proof of Theorem~\ref{theo:approximation_yt}}\label{appendix:proof_theo_approximation_yt}
\begin{proof}
The two observation equations \eqref{eq:state_space1} and \eqref{eq:proposed_ssm1} differ only in the latent state, therefore
\begin{align*}
\bm y(\bm s, t)-\bm y^{R}(\bm s, t)&=\bm X_t'\bm \beta+\bm W\bm z(\bm s, t)+\bm \varepsilon(\bm s,t)-\bm X_t'\bm \beta+\bm W\bm z^R(\bm s, t)+\bm \varepsilon(\bm s,t) \\
&= \bm W(\bm z(\bm s, t)-\bm z^R(\bm s, t)),   
\end{align*}
so the mean–square bias \(\mathbb{E}\bigl[\lVert \bm y_t-\bm y_t^{R}\rVert^2\bigr] \) is controlled by the mean–square error of the state approximation:
\begin{align}\label{eq:bound_y_z}
\mathbb{E}\bigl[\lVert \bm y_t-\bm y_t^{R}\rVert^2_2\bigr]\le C \cdot \mathbb{E}\bigl[\lVert \bm z_t-\bm z_t^{R}\rVert^2_2\bigr],   
\end{align}
where $C$ is a constant that depends on $\|W\|$. The inequality \eqref{eq:consistency} implies
$$
\mathbb{E}\bigl[\lVert \eta_i^{R}-\eta_i\rVert_{\mathcal{H}^{-1}}^{2}\bigr]\;\le\;c\,h^{2},
$$
for each component $i=1,\dots,q$ of $\bm \eta$ where $\lVert\cdot\rVert_{\mathcal{H}^{-1}}$ is the dual norm of $\mathcal H^{1}$.
On the bounded domain $\Omega$, the inclusion $L^2(\Omega)\subset \mathcal{H}^{-1}(\Omega)$ leads to 
$$
\lVert g\rVert_{L^{2}}\;\le\;C_{\mathrm{emb}}\;\lVert g\rVert_{H^{-1}}
\quad(\text{for every }g\in H^{-1}\ \text{with zero mean}), 
$$
that applied to $g=\bm \eta_t^{R}-\bm \eta_t$ and taking expectation leads to 
$$
\mathbb{E}\bigl[\lVert \bm \eta^{R}-\bm \eta\rVert_{L^{2}}^{2}\bigr]
\;\le\;C_{\mathrm{emb}}^{2}\,\sum_{i=1}^q\mathbb{E}\bigl[\lVert  \eta_i^{R}-\eta_i\rVert_{H^{-1}}^{2}\bigr]
\;\le\;q\,C_{\mathrm{emb}}^{2}\,c\,h^{2}.
$$
As we are interested in a general time $t = 1,\dots, T$, the AR(1) equations \eqref{eq:state_space2} and \eqref{eq:proposed_ssm2} yield
\begin{align*}
\|\bm z_t-\bm z_t^{R}\|_2
      &\le \varphi \|\bm z_{t-1}-\bm z_{t-1}^{R}\|_2 + \|\bm \eta_t- \bm \eta_t^R\|_2,\\
      &\le  \varphi \|\bm z_{t-1}-\bm z_{t-1}^{R}\|_2 + q\, C_{\text{emb}}^2\, c\, h^2    
\end{align*}
with $C$ being a constant $\varphi = \max_i |f_i|<1$. Iterating backwards in time, we obtain 
\begin{align}\label{eq:bound_z_z0}
\mathbb{E}\bigl[\lVert \bm z_t^{R}-\bm z_t\rVert_{L^{2}}^{2}\bigr]
      \le \sum_{r = 0}^t \varphi^{2r}\, q\, C_{\text{emb}}^2\, c\, h^2  = \frac{1-\varphi^{2(t+1)}}{1-\varphi^2}\,  q\, C_{\text{emb}}^2\, c\, h^2
\end{align}
Equations \eqref{eq:bound_y_z} and \eqref{eq:bound_z_z0} yield the result. \hfill$\square$
\end{proof}


\section{Proof of Theorem~\ref{theo:parameter_bounds}}\label{appendix:parameter_bound}
Let $\bm y(\bm s, t)$ distributed as specified in SSM \eqref{eq:state_space1}--\eqref{eq:state_space2} with set of parameter $\Pi$. We are interested in studying the bias on the estimation of parameters when, for a fixed $R$ we use the LR-SSM in the estimation. Let $\bm \pi$ the vector $(\bm \beta ', \bm \sigma',\bm f', k_1,\dots, k_q)'$. Denote by 
$$
\mathcal P_{\bm \pi}\equiv\mathcal N\bigl(\bm X'(\bm s, t)\bm \beta,\bm \Sigma_y(\bm \pi)\bigr),\qquad 
\bm \Sigma_y(\bm \pi)=\bm W\bm \Sigma_z(\bm \pi)\bm W'+ \bm \Sigma,
$$
the law of  $\bm y(\bm s,t)$, with
$\bm \Sigma=\operatorname{diag}(\bm \sigma^{2})$ and latent covariance $\bm \Sigma_z(\bm \pi)$. Replacing $\bm \eta(\bm s,t)$ by its FEM-approximation $\bm \Psi_R(\bm s)\bm \eta(\mathcal{V}_R,t)$ yields a low-rank covariance
$$
\bm \Sigma_y^{(R)}(\bm \pi)=\bm W\bm \Sigma_{z,R}(\bm \pi)\bm W'+\bm\Sigma .
$$
Define the covariance perturbation
$$
\bm \Delta_R(\bm \pi):=\bm \Sigma_z(\bm \pi)-\bm \Sigma_{z,R}(\bm \pi).
$$
By the FEM approximation, see equation (11) of \cite{lindgren2011explicit}, the Galerkin error is bounded as
\begin{align}\label{eq:bound_h}
\bigl\|\Delta_R(\bm \pi)\bigr\| \le C\,h,
\end{align}
were $h$ is the diameter of the largest circle that can be inscribed in a triangle in the triangulation, and $C$ is a constant. Because the LR-SSM is misspecified, the likelihood based on \eqref{eq:proposed_ssm1}--\eqref{eq:proposed_ssm2} concentrates around the minimiser of the Kullback–Leibler divergence 
\begin{align}\label{eq:KL2}
\bm \pi_R^{\star}:=
\arg\min_{\bm\vartheta\in\Theta_R}\operatorname{KL}\!\bigl(\mathcal P_{\bm \pi}\,\|\,\mathcal P^{(R)}_{\bm\vartheta}\bigr)    
\end{align}
where $\mathcal P^{(R)}_{\bm\vartheta}\equiv\mathcal N(\bm X(\bm s, t)'\bm \beta,\bm \Sigma_y^{(R)}(\bm\vartheta))$.
For Gaussian measures with the same mean function, the Kullback–Leibler divergence takes the form
\begin{align}\label{eq:KL}
    \operatorname{KL}(\bm \pi,\bm\vartheta)
   =\tfrac12\left\{
        \operatorname{tr}\!\bigl[\bm\Sigma_y^{(R)}(\bm\vartheta)^{-1}\bm\Sigma_y(\bm \pi)\bigr]
       -\log\det\!\bigl[\bm\Sigma_y^{(R)}(\bm\vartheta)^{-1}\bm\Sigma_y(\bm \pi)\bigr]-p
   \right\}.
\end{align}
Setting the gradient to zero yields, after vectorising ($\operatorname{vec}$) the matrices,
$$
0 =\bm J'\bm 
     \Sigma_y^{-1}(\bm\pi)\otimes\bm \Sigma_y^{-1}(\bm\pi)\;
     \operatorname{vec}\bm \Delta_R(\bm \pi)
   + \;\bm H\,(\bm \pi_R^{\star}-\bm \pi)
   + o\!\left(\|\Delta_R(\bm \pi)\|\right),
$$
where $\bm J = \nabla_{\bm\vartheta}\operatorname{vec}\bm \Sigma_y^{(R)}(\bm \pi)\bigl|_{\bm\vartheta=\bm \pi}$ and $\bm H$ is the Fisher-information matrix of the correct model, which is positive definite.  Solving the equation, we obtain the exact first-order bias
\begin{align}\label{eq:zeroKL}
\bm \pi_R^{\star}-\bm \pi
   = -\bm H^{-1}
       \bm J'
       \bigl(\bm \Sigma_y^{-1}(\bm\pi)\otimes\bm \Sigma_y^{-1}(\bm\pi)\bigr)
       \operatorname{vec}\bm \Delta_R(\bm \pi)
   \;+\;o\!\bigl(\|\bm \Delta_R(\bm \pi)\|\bigr)
\end{align}
Combining \eqref{eq:zeroKL} with the covariance bound \eqref{eq:bound_h} delivers the rate
$$
\bigl\|\bm \pi_R^{\star}-\bm \pi\bigr\| = \mathcal O(h) .
$$
Note that, the term \eqref{eq:KL} does not involve the mean of the processes, which implies the parameter $\bm \beta$ has no bias.  \hfill$\square$

\section{EM Algorithm}\label{appendix:EM}
We are interested in estimating the set parameter $\Pi$ based on the observed data. We recall here the notation. Denote by $\bm{y}_t$ the $(m_t = \sum_{j=1}^p m_{j,t})$-dimensional vector
\[
\bm{y}_t = \left(\bm{y}_1(\mathcal{S}_{1,t},t)', \bm{y}_2(\mathcal{S}_{2,t},t)', \dots, \bm{y}_p(\mathcal{S}_{p,t},t)'\right)',
\]
where $\mathcal{S}_{j,t} = \{\bm{s}_{i,t}\}_{i=1}^{m_{j,t}}$ is the set of locations where the process $y_j(\cdot, t)$ has been observed, and
\[
\bm y_{j,t} = \bm{y}_j(\mathcal{S}_{j,t},t) = \left(y_j(\bm{s}_{1,t},t), \dots, y_j(\bm{s}_{m_{j,t},t},t)\right)'.
\]
Let $\bm{y}_{1:T}$ denote the vector $(\bm{y}_1', \dots, \bm{y}_T')'$.

An analogous notation is used for $\bm{z}_{1:T} = (\bm{z}_1', \dots, \bm{z}_T')'$, where
\[
\bm{z}_t = \left(\bm{z}_1(\mathcal{V}_R, t)', \dots, \bm{z}_q(\mathcal{V}_R, t)'\right)',
\quad \text{with} \quad
\bm{z}_{j,t} = \bm{z}_j(\mathcal{V}_R, t) = \left(z_j(\bm{r}_1, t), \dots, z_j(\bm{r}_R, t)\right)'.
\]
Moreover, let $\bm{X}_{1:T} = [\bm{X}_1; \dots; \bm{X}_T]$, where each $\bm{X}_t$ is a $b \times m_t$ matrix, with $b = \sum_{j=1}^p b_j$, defined as
\[
\bm{X}_t = \bm{X}_{1,t} \oplus \bm{X}_{2,t} \oplus \dots \oplus \bm{X}_{p,t}.
\]
Each component matrix $\bm{X}_{j,t}$ is constructed as
\[
\bm{X}_{j,t} = [\bm{X}_j(\bm{s}_{1,t}, t),\dots,\bm{X}_j(\bm{s}_{m_{j,t}, t}, t)],
\]
with $\bm{X}_{j,t}$ being a $b_j \times m_{j,t}$ matrix, and each column vector $\bm{X}_j(\bm{s}, t) \in \mathbb{R}^{b_j}$ representing the covariate values associated with the $j$-th component of $\bm{y}(\bm{s}, t)$ at location $\bm{s}$ and time $t$.

We assume that the mesh $\mathcal{G}_R = (\mathcal{V}_R, \mathcal{E}_R)$, on which the latent process in the state-space model \eqref{eq:proposed_ssm1}–\eqref{eq:proposed_ssm2} is defined, is given. The procedure for selecting this mesh is discussed in Section~\ref{sec:triangulation}.

Let $L(\Pi; \bm y_{1:T},\bm X_{1:T}, \bm z_{1:T}) = p(\bm y_{1:T}, \bm z_{1:T}|\bm X_{1:T};\Pi)$ be the full likelihood function. The MLE of $\Pi$ is obtained by maximising the marginal likelihood of the observed data likelihood  
\begin{equation}
    p(\bm{y}_{1:T} | \bm X_{1:T}; \Pi) = \int p(\bm{y}_{1:T} | \bm X_{1:T},\bm{z}_{1:T}, \Pi) p(\bm{z}_{1:T} | \Pi) d\bm{z}_{1:T}.
\end{equation}
However, this quantity is intractable since $\bm z_{1:T}$ is unobserved and its distribution is unknown before attaining $\Pi$. The EM algorithm seeks to find the ML estimate of the marginal likelihood by iteratively applying these two steps:
\begin{itemize}
    \item[\textbf{E-step}]  Define \( Q(\Pi \mid \Pi^{(k)}) \) as the expected value of the log-likelihood function of \( \Pi \), with respect to the current conditional distribution of \( \bm z_{1:T}\) given \( (\bm y_{1:T}, \bm X_{1:T})  \) and the current estimates of the parameters \( \Pi^{(k)}\):  \[
 Q(\Pi \mid \Pi^{(k)}) = \mathbb{E}_{\bm z_{1:T} \sim p(\cdot \mid \bm y_{1:T}, \bm X_{1:T};\Pi^{(k)})} [\log p(\bm y_{1:T}\mid \bm X_{1:T},\bm{z}_{1:T};\Pi)].
\]
\item[\textbf{M-step}]   Obtain the next value of $\Pi$, namely $\Pi^{(k+1)}$ as the maximiser of $Q(\Pi|\Pi^{(k)}):$
\begin{equation}\label{eq:maximization}
\Pi^{(k+1)} = \arg \max_{\Pi\in \mathcal{P} } Q(\Pi \mid \Pi^{(k)}),    
\end{equation}
where the maximisation is done on a suitable compact set $\mathcal{P}$ which contains $\Pi$ as interior point. 
\end{itemize}
The computation of the conditional distribution of the latent states \(\bm{z}_{1:T}\) given the observed data and the current value of the parameter $\Pi^{(i)}$ can be efficiently obtained using the Kalman filter and smoother, and the formulas are given in the Appendix~\ref{appendix:EM} together with the formulas to obtain $\arg \max$ in \eqref{eq:maximization}. Several stopping criteria can be considered, including a maximum number of iterations, a tolerance on the norm of the difference \( \Pi^{(k)} - \Pi^{(k+1)} \), or a tolerance on the percentage change in the likelihood. Suppose that the algorithm stops after $K$ iterations, then the value $\hat{\Pi} = \Pi^{(K)}$ corresponds to the proposed estimator of $\Pi$.

\paragraph*{Close formulas} We now obtain close formulas of the parameters \( \{\bm{\beta}, \bm{\sigma}^2, \bm{f}, \bm w\}\) and express the objective function that we need to numerically minimise to obtain an estimate of the value of $\kappa_i$ for $i = 1,\dots,q$. First, we write explicitly the full log-likelihood $\ell(\Pi; \bm y_{1:T},\bm x_{1:T}, \bm z_{1:T}) = \log p(\bm y_{1:T}, \bm z_{1:T}|\bm x_{1:T};\Pi)$. Denote by $w_{ij}$ the element at row $i$ and column $j$ of the matrix $\bm W$ in \eqref{eq:proposed_ssm1} and  define the matrix $\bm \Psi_t^{\bm w}$ as 
\begin{align}\label{eq:matrix_loading}
\bm \Psi_t^{\bm w} = \begin{bmatrix}
     w_{11} \bm \Psi_{11,t} & \dots & w_{1q} \bm \Psi_{1q,t}\\
     \vdots & \ddots & \vdots \\
     w_{p1} \bm \Psi_{p1,t} & \dots & w_{pq}\bm  \Psi_{pq,t}
\end{bmatrix},
\end{align}
where \( \bm \Psi_{ij,t} \) is a known basis matrix that characterises the effect of the vector \( \bm{z}_j(\mathcal{V}_j,t) \) on \( \bm{y}_i(\mathcal{S}_{i,t}, t) \), that is $\bm \Psi_{ij,t} = [\bm \psi_R(\bm s_{1,t})';\dots;\bm \psi_R(\bm s_{m_{i,t},t})']$. 
Using the notation introduced in Section \ref{sec:estimation}, we can write the SSM \eqref{eq:proposed_ssm1}-\eqref{eq:proposed_ssm2} in matrix form as
\begin{align*}
    \bm y_t &= \bm X_t'\bm \beta + \bm \Psi_t^{\bm w} \bm z_t + \bm \varepsilon_t\\
    \bm{z}_t&= \bm F \bm{z}_{t-1} + \bm{\eta}_t,  
\end{align*}
where $\bm \varepsilon_t = (\bm \varepsilon_1(\mathcal{S}_{1,t},t)', \dots, \bm \varepsilon_p(\mathcal{S}_{p,t},t))'$ with $\bm \varepsilon_i(\mathcal{S}_{i,t},t) = (\varepsilon_i(\bm s_{1,t},t), \dots, \varepsilon_i(\bm s_{m_{i,t},t},t))'$. Thus, the full log-likelihood $\ell (\Pi; \bm y_{1:T},\bm X_{1:T}, \bm z_{1:T})$ satisfies 
\begin{align*}
    -2\ell &(\Pi; \bm y_{1:T},\bm X_{1:T}, \bm z_{1:T}) = \log |\bm \Sigma_0|+(\bm z_0-\bm \mu_0)'\bm \Sigma_0^{-1}(\bm z_0-\bm \mu_0)\\
    &+ \sum_{t=1}^T \left(\log |\bm\Sigma_t| + \bm e_t'\bm\Sigma_t^{-1}\bm e_t \right)- T\log |\bm Q_{\bm \kappa}| +  \sum_{t=1}^T \left(\bm z_t-\bm F \bm z_{t-1}\right)' \bm Q_{\bm \kappa} \left(\bm z_t-\bm F \bm z_{t-1}\right),
\end{align*}
where $\bm e_t = \left(\bm y_t-\bm X'_t\bm\beta-\bm \Psi_t^{\bm w} \bm z_t\right)$, and $\bm \Sigma_t = (\sigma_1 \bm I_{m_1,t}) \oplus \dots \oplus (\sigma_p\bm I_{m_p,t}) $ is the covariance matrix of 
$\bm \varepsilon_t$, and $\bm{Q}_{\bm \kappa} = \bm{Q}_{\kappa_1} \oplus \dots \oplus \bm{Q}_{\kappa_q}$ is the precision matrix of the innovation term $\bm{z}_t - \bm{F} \bm{z}_{t-1}$, the matrix $\bm F = (f_1\bm I_{R_1})\oplus \dots\oplus (f_q\bm I_{R_q})$ corresponds to the transition matrix and $\bm \mu_0$ and $\bm \Sigma_0$ are the vector mean and covariance matrix of $\bm z_0$, respectively. 
Thus, the conditional expectation of the complete-data log-likelihood given the current estimate of $\Pi$ at pass $(k)$ of the EM algorithm takes the form
\begin{align}\label{eq:EM_Q}
    Q(\Pi|\Pi^{(k)}) &= \mathbb{E}_{\Pi^{(k)}}\left[-2\ell (\Pi; \bm y_{1:T},\bm x_{1:T}, \bm z_{1:T})\vert \bm y_{1:T},\bm X_{1:T}\right]
\end{align}
where \(\mathbb{E}_{\Pi^{(k)}}\) stands for the expected value considering the posterior distribution of the latent variables \(\bm z_{1:T}\) given the observed data \((\bm y_{1:T}, \bm X_{1:T})\) and the current parameter estimate \(\Pi^{(k)}\). 

Applying the conditional expectation, it follows that
\begin{align}\label{eq:EM_Q_MAX}
     &Q(\Pi|\Pi^{(k)})  = \log |\bm \Sigma_0|+\text{Tr}\left[ \bm \Sigma_0^{-1} \left( (\bm z_0^T-\bm \mu_0)(\bm z_0^T-\bm \mu_0)' + \bm P_0^T\right)\right] \nonumber\\
    &\,+ \sum_{t=1}^T \log |\bm\Sigma_t| + \sum_{t=1}^T \text{Tr}\left[ \bm\Sigma_t^{-1}\left(\left(\bm y_t-\bm X'_t\bm\beta-\bm \Psi_t^{\bm w} \bm z_t^T\right)
    \left(\bm y_t-\bm X'_t\bm\beta-\bm \Psi_t^{\bm w} \bm z_t^T\right)' + \bm \Psi_t^{\bm w}\bm P_t^T\bm \Psi^{\bm w}\right)\right] \nonumber\\
    & - T\log |\bm Q_{\bm \kappa}| + \text{Tr}\Big[\bm Q_k\sum_{t=1}^T(\bm z_t^T- \bm F\bm z_{t-1}^T)(\bm z_t^T- \bm F\bm z_{t-1}^T)'\Big] \nonumber\\
    &+\text{Tr}\Big[\bm Q_{\bm \kappa} \sum_{t=1}^T \big(
    \bm P_t^T + \bm F\bm P_{t-1}^{T}\bm F'- \bm F \bm P_{t,t-1}^T - (\bm P_{t,t-1}^T)' \bm F'  \big)\Big] 
\end{align}
where the quantities $\bm z_t^t$ and $\bm P_t^t$ for $t = 0,\dots, T$ are obtained sequentially using the Kalman Filter equations, whose forward pass for $t =,1,\dots, T$ consists of:
\begin{align*}
        \textbf{Prediction:}\quad  \bm{z}_t^{t-1} &= \bm{F}^{(k)} \bm{z}_{t-1}^{t-1}, \\
        \bm{P}_t^{t-1} &= \bm{F}^{(k)} \bm{P}_{t-1}^{t-1} \bm{F}^{(k)'} + \bm{Q}_{\bm{\kappa}^{(k)}}^{-1},\\
    \text{with}\quad  \bm{z}_0^0 &= \bm{\mu}_0^{(k)}, \quad \bm{P}_0^0 = \bm{\Sigma}_0^{(k)}.\\
 \textbf{Update (Filter):} \quad \bm{\epsilon}_t &= \bm{y}_t - \bm X_t'\bm \beta^{(k)}- \bm \Psi_t^{\bm{w}^{(k)}} \bm{z}_t^{t-1}, \\
        \bm{\Sigma}_{\bm{\epsilon}_t} &= \bm \Psi_t^{\bm{w}^{(k)}} \bm{P}_t^{t-1} (\bm \Psi_t^{\bm{w}})' + \bm{\Sigma}_t, \\
        \bm{K}_t &= \bm{P}_t^{t-1} (\bm \Psi_t^{\bm{w}^{(k)}})' (\bm{\Sigma}_{\bm{\epsilon}_t})^{-1}, \\
        \bm{z}_t^t &= \bm{z}_t^{t-1} + \bm{K}_t \bm{\epsilon}_t, \\
        \bm{P}_t^t &= \bm{P}_t^{t-1} - \bm{K}_t \bm \Psi_t^{\bm{w}^{(k)}} \bm{P}_t^{t-1}.
    \end{align*}
and where the quantities $\bm z_t^T$ and $\bm P_t^T$ for $t = 0,\dots, T$ are obtained sequentially using the Kalman Smoothing equations, whose backwards pass for $t =T,\dots, 1$ consists of:
\begin{align*} 
 \textbf{Smoothing:}\quad  \bm{z}_T^T &= \bm{z}_T^T, \quad \bm{P}_T^T = \bm{P}_T^T, \quad \text{(for } t = T \text{ only)} \\
\bm{J}_{t-1} &= \bm{P}_{t-1}^{t-1} \bm{F}^{(k)'} (\bm{P}_{t}^{t-1})^{-1}, \\
\bm{z}_{t-1}^T &= \bm{z}_{t-1}^{t-1} + \bm{J}_{t-1} (\bm{z}_{t}^T - \bm{z}_{t}^{t-1}), \\
\bm{P}_{t-1}^T &= \bm{P}_{t-1}^{t-1} + \bm{J}_{t-1} (\bm{P}_{t}^T - \bm{P}_{t}^{t-1}) \bm{J}_t'. 
\end{align*}
The Smoothed lag-one covariance $\bm P_{t,t-1}^T$ for $t = 1,\dots, T$ is also obtain sequentially for $t = T, \dots, 2$ by
\begin{align*}
     \textbf{Smoothed lag-one covariance:}\quad \bm P_{T,T-1}^T &= (\bm I-\bm K_T\bm \Psi_{\bm w^{(k)}}^T)\bm F \bm P_{T-1}^{T-1}, \quad \text{(for } t = T \text{ only)} \\
\bm P_{t,t-1}^T &= \bm P_{t,t-1} \bm J_{t-2}' + \bm J_{t-2} (\bm P_{t-1,t-1}^T- \bm F \bm P_{t-1}^{t-1})\bm J_{t-2}'.
\end{align*}

Now, we differentiate \( Q(\Pi | \Pi^{(k)}) \) from equation \eqref{eq:EM_Q_MAX} concerning \( \bm{\beta}, \bm{\sigma}, \bm{w}, \bm{f} \) to derive closed-form expressions for the values that minimise the function. Meanwhile, the objective function, which must be minimised numerically, provides an estimate for \( \kappa_i \) for \( i = 1, \dots, q \). 

Denote by $\bm e_i$ the $q-$dimensional vector with all entries equal to zero except for the $i-$th that is equal to 1. Moreover, for $i = 1,\dots, p$,  define \(\bm \Psi_{i,t}
=\bm \Psi_{i1,t}\oplus \dots\oplus \bm \Psi_{iq,t}\) that is a $(q m_{i,t})\times (qR)$ matrix.


We obtain:
\begin{align}\label{eq:close_formula_beta}
 \bm \beta = \left( \sum_{t=1}^T \bm X_t' (\bm\Sigma_t^{(k)})^{-1} \bm X_t \right)^{-1}
\left( \sum_{t=1}^T \bm X_t' (\bm\Sigma_t^{(k)})^{-1}  \left( \bm y_t - \bm \Psi_t^{\bm w^{(k)}} \bm z_t^T \right) \right)
\end{align}
For \( i = 1, \dots, p \), update \( (\sigma_i^2)^{(k+1)} \) as:
\begin{equation}\label{eq:close_formula_sigma}
\begin{aligned}
(\sigma_i^2)^{(k+1)} =\left(\sum_{t=1}^T m_{i,t}\right)^{-1}\sum_{t=1}^T \text{Tr} \Bigg\{ &\left( \bm{y}_{i,t} - \bm{X}_{i,t}' \bm{\beta}^{(k)} - \left[w_{i1}^{(k)} \bm \Psi_{i1}^t, \dots, w_{iq}^{(k)} \bm \Psi_{iq}^t\right] \bm{z}_t^T \right) \\
&\times \left( \bm{y}_{i,t} - \bm{X}_{i,t}' \bm{\beta}^{(k)}  - \left[w_{i1}^{(k)} \bm \Psi_{i1}^t, \dots, w_{iq}^{(k)} \bm \Psi_{iq}^t\right] \bm{z}_t^T \right)' \\
&+  \left[w_{i1}^{(k)} \bm \Psi_{i1}^t, \dots, w_{iq}^{(k)} \bm \Psi_{iq}^t\right] \bm{P}_t^T \left[w_{i1}^{(k)} \bm \Psi_{i1}^t, \dots, w_{iq}^{(k)} \bm \Psi_{iq}^t\right]' \Bigg\}.
\end{aligned}
\end{equation}
Continuing, we have 
\begin{align}\label{eq:close_formula_f}
    f_i &= \frac{\text{Tr}\left[\bm Q_{\kappa_i^{(k)}} \sum_{t=1}^T (\bm e_i' \otimes \bm I_R) (\bm P_{t,t-1}^T + \bm z_{t-1}^T\bm z_t^{T'} )(\bm e_i \otimes \bm I_R)  \right]}{\text{Tr}\left[\bm Q_{\kappa_i^{(k)}} \sum_{t=1}^T  (\bm e_i' \otimes \bm I_R)  (\bm P_{t-1}^T + \bm z_{t-1}^T\bm z_{t-1}^{T'} ) (\bm e_i\otimes \bm I_R) \right]}.
\end{align}
Moreover, 
\begin{align*}
    \bm \mu_0 = \bm z_0^T, \quad \text{and} \quad \bm \Sigma_0 = \bm P_{0}^T.
\end{align*}
Finally, to derive the closed-form expression for $\bm{w}$, proceed as follows. 

Let us consider the quantity in the likelihood involving the only $\bm w_i = (w_{i1},\dots, w_{iq})'$, that is:
\begin{align*}
   \sum_{t=1}^T \mathbb{E}\left[ \left \|\bm y(\mathcal{S}_{i,t},t)-\bm{X}^i(\mathcal{S}_{i,t}), t - \sum_{j=1}^qw_{ij} (\bm e_j'\otimes \bm I_{m_i,t})\bm \Psi_{i,t}\bm z_t\right\|^2\big \vert\bm y_{1:T}, \bm X_{1:T}\right],
\end{align*}
where $\|\cdot\|$ denotes the euclidean norm. Observe that
\begin{align*}
    \sum_{j=1}^qw_{ij} (\bm e_j'\otimes \bm I_{m_i,t})\bm \Psi_{i,t}\bm z_t = \left [(\bm e_1'\otimes \bm I_{m_i,t})\bm \Psi_{i,t}\bm z_t, \dots, (\bm e_j'\otimes \bm I_{m_i,t})\bm \Psi_{i,t}\bm z_t\right ] \bm w_i.
\end{align*}
Therefore, we have to minimise in $\bm w_i$ the quantity 
\begin{align}\label{eq:obj_w}
    &-2 \sum_{t=1}^T \left( \bm y(\mathcal{S}_{i,t},t)-\bm{X}^i(\mathcal{S}_{i,t}, t \right)'\mathbb{E}\left [(\bm e_1'\otimes \bm I_{m_i,t})\bm \Psi_{i,t}\bm z_t, \dots, (\bm e_j'\otimes \bm I_{m_i,t})\bm \Psi_{i,t}\bm z_t\right \vert\bm y_{1:T}, \bm X_{1:T}] \bm w_i \nonumber\\
    &\quad +\bm w_i' \mathbb{E}\Bigg[ \Big [(\bm e_1'\otimes \bm I_{m_i,t})\bm \Psi_{i,t}\bm z_t, \dots, (\bm e_j'\otimes \bm I_{m_i,t})\bm \Psi_{i,t}\bm z_t\Big ]' \nonumber\\
    &\quad\quad\quad\quad\quad\quad \times  \Big [(\bm e_1'\otimes \bm I_{m_i,t})\bm \Psi_{i,t}\bm z_t, \dots, (\bm e_j'\otimes \bm I_{m_i,t})\bm \Psi_{i,t}\bm z_t\Big ]  \Bigg\vert\bm y_{1:T}, \bm X_{1:T}\Bigg] \bm w_i.
\end{align}
The first term in \eqref{eq:obj_w} is equivalent to $-2\bm g_i'\bm w_i$, where 
\begin{align}
    \bm g_i' = \sum_{t=1}^T\left( \bm y(\mathcal{S}_{i,t},t)-\bm{X}^i(\mathcal{S}_{i,t}, t )\bm \beta^{(k)}\right)'\left [(\bm e_1'\otimes \bm I_{m_i,t})\bm \Psi_{i,t}\bm z_t^T, \dots, (\bm e_q'\otimes \bm I_{m_i,t})\bm \Psi_{i,t}\bm z_t^T\right].
\end{align}
The second term in \eqref{eq:obj_w} is equivalent to $  \bm w_i'\bm R_i \bm w_i$, where $\bm R_i = [r_{i,jk}]_{j,k=1}^q$ is a $q\times q$ matrix with  
\begin{align*}
  r_{i,jk} &= \sum_{t=1}^T\mathbb{E}\left[\bm z_t^{'} \bm \Psi_{i,t}' (\bm e_j'\otimes \bm I_{m_i,t})' (\bm e_k'\otimes \bm I_{m_i,t})\bm \Psi_{i,t}\bm z_t^{}\vert\bm y_{1:T}, \bm X_{1:T}\right]\\
  &=\sum_{t=1}^T\bm z_t^{T'} \bm \Psi_{i,t}' (\bm e_j'\otimes \bm I_{m_i,t})' (\bm e_k'\otimes \bm I_{m_i,t})\bm \Psi_{i,t}\bm z_t^{T} + \operatorname{Tr}\left( \bm \Psi_{i,t}'(\bm e_j'\otimes \bm I_{m_i,t})' (\bm e_k'\otimes \bm I_{m_i,t})\bm \Psi_{i,t}\bm P_t^T\right)\\
  &= \,
   \sum_{t=1}^{T}
      \operatorname{Tr}\left(
         \bm\Psi_{i,t}'\left((\bm e_j\bm e_k') \otimes \bm I_{m_i,t}\right)\bm\Psi_{i,t}
       \left( \bm z_t^T\bm z_t^{T'}+\bm P_t^{T}\right)\right).
\end{align*}

Therefore, we have to minimise 
\begin{align}
 -2\bm g_i'\bm w_i +\bm w_i'\bm R_i \bm w_i,
\end{align}
and therefore
\begin{align}
    \bm w_i^{(k+1)} = \bm R_i^{-1} \bm g_i
\end{align}

For the updating the value of $\kappa_i$ we numerically minimise the the function $g_i^{(k)}:\mathbb{R}_+\to \mathbb{R}_+$ defined as 
\begin{align*}
    g_i(\kappa_i) &= - T\log |\bm Q_{\bm \kappa_i}| + \text{Tr}\Big[\bm Q_{\bm \kappa_i} \sum_{t=1}^T(\bm e_i'\otimes \bm I_R)(\bm z_t^T- \bm F^{(k)}\bm z_{t-1}^T)(\bm z_t^T- \bm F^{(k)}\bm z_{t-1}^T)'(\bm e_i\otimes \bm I_R)\Big] \\
    &\quad + \text{Tr}\Big[\bm Q_{\bm \kappa_i} \sum_{t=1}^T (\bm e_i'\otimes \bm I_R)\big(
    \bm P_t^T + \bm F^{(k)}\bm P_{t-1}^{T}\bm F^{(k)'} -\bm F^{(k)} \bm P_{t,t-1}^T - (\bm P_{t,t-1}^T)' \bm F^{(k)'} \big)(\bm e_i\otimes \bm I_R)\Big],
\end{align*}
therefore
$$
\kappa_i^{(k+1)} = \arg\min_{\kappa_i}  g_i^{(k)}(\kappa_i).
$$


\section{Boundary Effects} \label{appendix:boundary}
In this section, we recall the study of the impact of the Neumann boundary conditions on the approximation of the error term $\bm \eta(\bm s, t)$ as in \cite{lindgren2011explicit}. In particular, consider the SPDE:
\begin{equation}\label{eq:spde_full}
\begin{split}
    (\kappa^2 - \Delta)^{\alpha/2} x(\bm{u}) &= \mathcal{W}(\bm{u}), \quad \bm{u} \in \Omega, \\
    \partial_n (\kappa^2 - \Delta)^j x(\bm{u}) &= 0, \quad \bm{u} \in \partial \Omega, \quad j = 0, 1, \dots, \left\lfloor \frac{\alpha - 1}{2} \right\rfloor,
\end{split}
\end{equation}
where \eqref{eq:spde_full} uses the same notation as in Appendix~\ref{appendix:weak_approxiation}.  In the one-dimensional case, the following result characterises the covariance structure induced by these boundary conditions. The result naturally extends to higher-dimensional domains that are generalised rectangles in \( \mathbb{R}^d \).
\begin{theorem}
Let \( x \) be the solution to \eqref{eq:spde_full} on the domain \( \Omega = [0, L] \), with integer order \( \alpha > 0 \). Then the covariance function of \( x \) is given by
\[
\text{cov} \{ x(u), x(v) \} = \sum_{k=-\infty}^{\infty} \left\{ r_M(u, v - 2kL) + r_M(u, 2kL - v) \right\},
\]
where \( r_M \) denotes the Matérn covariance function defined on the real line \( \mathbb{R} \).
\end{theorem}
This result reveals that the covariance can be interpreted as a sum over reflections (or foldings) of the Matérn covariance across the domain boundaries. When the effective range of the Matérn covariance is small compared to the domain size \( L \), an accurate approximation is obtained by retaining only the three dominant terms:
\begin{align*}
    \text{cov} \{ x(u), x(v) \} &\approx r_M(u, v) + r_M(u, -v) + r_M(u, 2L - v),  \\
    &= r_M(0, v - u) + r_M(0, v + u) + r_M(0, 2L - (v + u)). 
\end{align*}
Furthermore, away from the boundaries—specifically, at distances greater than twice the range parameter from the domain edges—the folded covariance becomes nearly indistinguishable from the stationary Matérn covariance on \( \mathbb{R} \). This argument justifies the proposed construction of the precision matrix $\bm Q_{k_i}$ in Section~\ref{sec:boundary}.

\bibliographystyle{apalike}
\bibliography{references}

\end{document}